\documentclass[11pt,a4paper]{article}
\usepackage{url}

\usepackage{amsmath,amsfonts,natbib,graphicx,psfrag,amssymb}
\usepackage{enumerate,enumitem,verbatim,xspace,comment}
\usepackage{color}
\def\hang{\hangindent=8truemm\hangafter=1\smallskip\noindent}

\renewcommand{\mid}{|}
\newcommand{\rjb}[1]{\textcolor{red}{#1}}

\newcommand{\Ga}{\text{Ga}}
\newcommand{\Po}{\text{Po}}
\newcommand{\LN}{\text{LN}}
\renewcommand{\vec}[1]{\boldsymbol{#1}}

\newcommand{\xvec}{\vec{x}}
\newcommand{\cvec}{\vec{c}}

\newcommand{\bvec}{\vec{b}}
\newcommand{\dvec}{\vec{d}}

\newcommand{\thetavec}{\vec{\theta}}
\newcommand{\deltavec}{\vec{\delta}}

\newcommand{\lambdavec}{\vec{\lambda}}
\newcommand{\kappavec}{\vec{\kappa}}
\newcommand{\alphavec}{\vec{\alpha}}

\newcommand{\nuvec}{\vec{\nu}}
\newcommand{\xivec}{\vec{\xi}}

\newcommand{\omegavec}{\vec{\omega}}
\newcommand{\etavec}{\vec{\eta}}
\newcommand{\pivec}{\vec{\pi}}
\newcommand{\diag}{\textrm{diag}}
\newcommand{\indep}{\overset{indep}{\sim }}
\renewcommand{\d}{\phantom{2}}
\newcommand{\numtestcricketers}{2855\xspace}
\newcommand{\finaltestnum}{2269}
\newcommand{\finaltestdate}{August 2017}
\newcommand{\finaltestyear}{2017\xspace}
\newcommand{\dataspan}{141\xspace} 
\newcommand{\numnonzerodelta}{140\xspace}

\title{On the ranking of Test match batsmen}

\author{Richard J. Boys$^1$\footnote{richard.boys@ncl.ac.uk}, Peter M. Philipson$^2$}
\date{\small $^1$ School of Mathematics, Statistics and Physics,
  Newcastle University, UK\\$^{2}$Mathematics, Physics and Electrical
  Engineering, Northumbria University, UK\\}

\begin{document}
\maketitle 

\begin{abstract}
  Ranking sportsmen whose careers took place in different eras is
  often a contentious issue and the topic of much debate. In this
  paper we focus on cricket and examine what conclusions may be drawn
  about the ranking of Test batsmen using data on batting scores from
  the first Test in 1877 onwards. The overlapping nature of playing
  careers is exploited to form a bridge from past to present so that
  all players can be compared simultaneously, rather than just
  relative to their contemporaries. The natural variation in runs
  scored by a batsman is modelled by an additive log-linear model with
  year, age and cricket-specific components used to extract the innate
  ability of an individual cricketer.  Incomplete innings are handled
  via censoring and a zero-inflated component is incorporated into the
  model to allow for an excess of frailty at the start of an innings.
  The innings-by-innings variation of runs scored by each batsman
  leads to uncertainty in their ranking position. A Bayesian approach
  is used to fit the model and realisations from the posterior
  distribution are obtained by deploying a Markov Chain Monte Carlo
  algorithm. Posterior summaries of innate player ability are then
  used to assess uncertainty in ranking position and this is
  contrasted with rankings determined via the posterior mean runs
  scored. Posterior predictive checks show that the model provides a
  reasonably accurate description of runs scored.
\end{abstract}

\noindent\text{Keywords:} Censoring; Overdispersion, Poisson random effects; Zero-inflation

\section{Introduction}
There is a great deal of discussion in many sports, from the experts through to the fans, about who is the `greatest'. Discussions often conclude with the notion that it is impossible to obtain definitive answers. In many cases the game played out in the modern day, in front of the massed media with large teams of supporting staff dedicated to nutrition, fitness and psychology, bears little or no relation to the backdrop at the genesis of the sport. The richness of data now available however suggests that there may be merit in a sophisticated statistical approach to the problem.

The analysis of sports data has undergone something of a boom in recent years with statisticians and data analysts at the forefront. In baseball, for example, sabermetrics has become an accepted term for the use of in-game statistical analysis \citep{RBaseball}, and there is an increasing trend of sports science and data analysis being routinely performed by major sports organisations across the globe.

In this paper we focus on the sport of cricket and look at the
performance of Test match batsmen. Cricket is a bat-and-ball game
played between two teams of eleven players each on a cricket field, at
the centre of which is a rectangular 22-yard-long pitch with a target
called the wicket (a set of three wooden stumps topped by two bails)
at each end. Each phase of play is called an innings during which one
team bats, attempting to score as many runs as possible, whilst their
opponents field. In Test matches the teams have two innings apiece
and, when the first innings ends, the teams swap roles for the next
innings.  This sequence can only be altered by the team batting second
being made to `follow-on' after scoring significantly fewer runs than
the team batting first. Except in matches which result in a draw, the
winning team is the one that scores the most runs, including any
extras gained.  Individual players start their innings with zero runs
and accumulate runs as play progresses, leading to a final score which
is a non-negative count. The highest individual score in Test cricket
is 400 runs and the average score is around 30 runs. Smaller scores
are more likely than larger scores as the aim of the opposition is to
bowl out each batsman as quickly as possible and at the cost of as few
runs as possible.

The earliest work on the statistical modelling of cricket scores was
undertaken by \cite{Elderton1945} and \cite{Wood1945} who considered
modelling samples of individual first-class cricket scores from both
Test matches and the County Championship (the domestic first-class
cricket competition in England and Wales, sitting one level below Test
cricket) as a geometric progression and found evidence of a reasonable
fit, although Wood commented that the `series show discrepancies at
each end, and particularly at the commencement' due to a larger than
expected number of scores of zero, or ducks in cricketing parlance.
Incomplete (`not-out') scores were assumed to continue at the start of
the next innings in the former (acknowledged as a `pleasant fiction'
by the author) and treated as complete innings by the latter. Later
\cite{Pollard77} investigated the distribution of runs scored by teams
in county cricket and found the negative binomial distribution to
offer a good fit.  \cite{Scarf2011} confirmed this finding for runs
scored in both batting partnerships and team innings in Test cricket.

\cite{Kimber} considered the merits of the geometric distribution for
samples of individual cricket scores from Test and first class
matches, including Australia's domestic Sheffield Shield competition,
along with one day internationals, concluding that `there was little
evidence against the \ldots~model in the upper tail' but rejecting its
validity for low scores, mainly due to the excess of ducks in the
data. Their work focused on an alternative batting average measure
using a non-parametric product-limit estimation approach. Some of
these points will be revisited later.  They also looked at the
independence of cricket scores for a batsman and found `no major
evidence of autocorrelation' via a point process approach, surmising
that `it is quite reasonable \ldots~to treat scores as if they were
independent and identically distributed
observations'. \cite{Durbach2007} later concluded that batting scores
can be considered to come from a random sequence based on a study of
sixteen Test match batsmen. We note that studies in other sports of
lack of independence of points- or run-scoring, sometimes referred to
as the `hot hand', have largely concluded that there is little
evidence to support the notion of `form' \citep{Gilovich85,
  Tversky89}.

Published work in sports statistics covers a wide range of sports.
Initially much of this work centred around the analysis of football
and baseball, and focussed on predicting future outcomes but now
increasingly looks at gains that might be made using an optimal
strategy. The most famous model-based method used in cricket today is,
of course, the Duckworth--Lewis--Stern formula \citep{DuckworthL1998,
  DuckworthL2004, Stern2009} for interrupted one-day cricket matches,
with subsequent modification by, for example, \cite{McHaleA2013}.
Other work such as \cite{SilvaMS2015} looks at the effect of powerplay
in such matches. In this paper the focus is instead on comparing past
and current players, an area where relatively little research has been
done \citep{Rohde2011, Radicchi2011, Baker2014}, and study Test
cricketers in particular.  The innate ability of each player is
modelled by taking into account the heterogeneous effect of ageing on
sporting performance, any year effects which act as a surrogate for
changes to the game that may have made it easier or more difficult in
certain eras, home advantage and some cricket-specific components.
\cite{BRL} considered how to compare players from different eras in
three, predominately US-based, sports: baseball, golf and ice hockey.
Their argument, which is adopted here, is that comparisons between
modern-day players and players from bygone eras are possible by
considering the overlap in playing careers: modern players at the
start of their careers will have played against older players at the
end of their careers, which started much earlier, and these older
players would, in their youth, have played against players from
earlier eras once more. In such a way a bridge from present to past is
formed.

The paper is structured as follows. The data are described in
Section~\ref{sec:data}. The model description in
Section~\ref{sec:model} begins by outlining an initial model before
introducing modifications to handle some nuances of cricket batting
data. Sections~\ref{sec:prior} and~\ref{sec:posterior} detail the
prior and posterior distributions respectively along with the MCMC
algorithm. Section~\ref{sec:results} describes some of the results
such as the posterior mean of player ability, and a ranking by this
measure, and summaries of the posterior distribution of player
rankings. The paper concludes with some discussion and avenues for
future work in Section~\ref{sec:discussion}.



\section{The data}
\label{sec:data}
Cricket is a highly data driven sport, perhaps more than any other
with the exception of baseball. Players' entire careers are typically
judged by a one-number summary: their average. There is a large amount
of data, typically in the form of scorecards, available for all
formats of cricket at both international, domestic and even regional
level. For some players there is even ball-by-ball data recorded (the
Association of Cricket Statisticians and Historians have these data
for Sir Jack Hobbs) although such a level of granularity is not
generally available and so is not considered further here.

The data used in this paper consists of individual innings by all Test
match cricketers ($n=\numtestcricketers$) from the first Test played
in 1877 up to Test \finaltestnum, in \finaltestdate. There are
currently ten Test playing countries and many more Test matches are
played today than at the time of the first Test. Indeed for the first
twelve years the combatants were exclusively England and Australia. A
demonstration of the growth of Test match cricket is given in
Table~\ref{Table:TestMatchTimeline}.  We note that, in contrast to the
standard presentation of historical batting averages such as at
\texttt{http://stats.espncricinfo.com/ci/content/records/282910.html},
we include all batsmen irrespective of the number of career innings
played. However, in keeping with other lists, we do not include World
Series Cricket matches as these matches are not considered official
Test matches by the International Cricket Council (ICC).
\begin{table}
\caption{\label{Table:TestMatchTimeline}Timeline of Test match milestones}
\centering
\fbox{%
\begin{tabular}{l c c}
\hline
Test Match \#&  Year& Elapsed years\\
\hline
1& 1877&  \\
100& 1908& 31 \\
500& 1960& 52 \\
1000& 1984& 24 \\
2000& 2011& 27 \\
\hline
\end{tabular}}
\end{table}

In addition to runs scored, the data contain other useful information
such as the venue, the opposing team and whether or not the batsman's
innings is incomplete (this can happen for a variety of reasons; see
Section~\ref{Subsec:censoring}). Thus we can determine whether a Test
match is played at `home' and investigate the extent of any home
advantage. Note that there have been twenty-nine Tests played at
neutral venues and we class these as away matches for both teams. The
data also include the match innings index which is potentially
important as (generally) the conditions in the final innings of a Test
match are at their worst for batting and the pressure, due to the game
situation, is often at its highest -- it is an axiom of cricket that
batting last is difficult.

Some aspects of the game that have changed over time are not
explicitly recorded in the data: at one stage Tests were `timeless',
continuing until a result was achieved; the number of balls in an over
has varied between four, five, six and eight; pitches were uncovered
and left exposed to the elements up to around 1960; the introduction
of limited-overs international cricket in 1971 along with the recent
advent of Twenty20 cricket in 2003 and the abolition of `amateur'
status in 1962. Together these aspects may well have affected the
performance of batsmen, particularly possible changes to pitches and
changes to the game dynamic induced by the shorter formats. We will
consider these on annual and decade scales respectively.

\begin{figure}[t]
\centering
\includegraphics[width=0.7\textwidth]{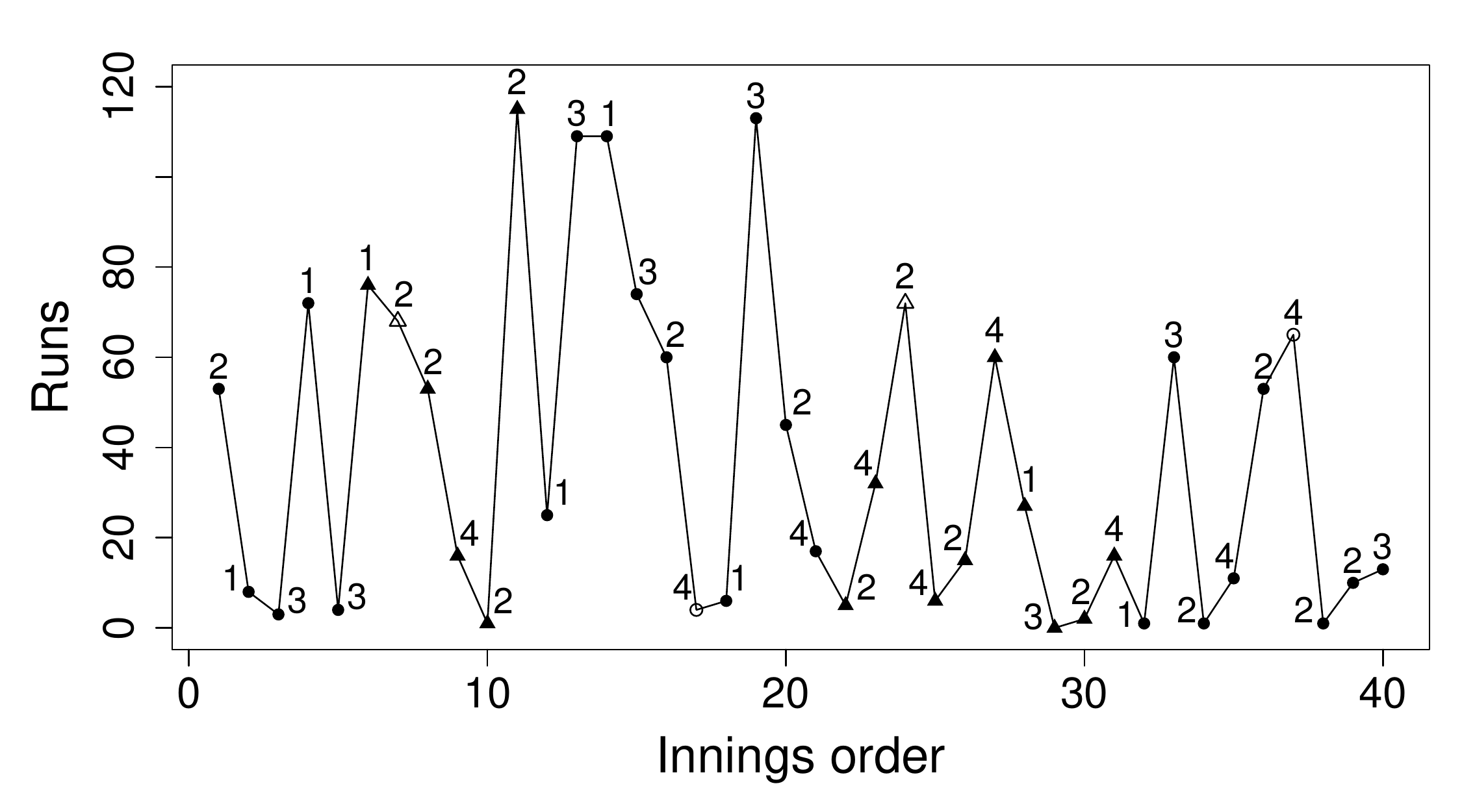}
\caption{Innings played by Ian Bell against Australia between 2009 and
  2015. Triangles -- away matches; open symbols -- not outs; numbers
  -- innings number.}
\label{fig:IanBell}
\end{figure}

A typical profile of batting scores is given in
Fig.~\ref{fig:IanBell}. This plot shows the runs scored in the Test
match innings of England batsman Ian Bell against Australia between
2009 and \finaltestyear. The innings are shown in sequential order and
away matches and not outs are indicated by the plotting symbol. Note
that, although Bell was in the England side throughout this period, he
did not bat in many innings. This feature is typical and can be due to
many factors such as big wins where the follow-on was deployed and the
winning team did not need to bat a second time, or if the match is
drawn due to bad weather or running out of time.  The figure
highlights the capricious nature of batting and suggests that, while
year, ageing and game-specific effects may affect run-scoring on an
overall level, the innings-by-innings variation is considerable.

\section{The model}
\label{sec:model}
Runs scored in an innings are counts and a natural starting point is
to consider modelling them via the Poisson distribution, with
\[
X_{ijk}|\lambda_{ijk}\indep\Po(\lambda_{ijk}), \qquad i=1,\ldots,\numtestcricketers; \quad
j=1,\ldots,n_i; \quad k=1,\ldots, n_{ij},
\]
where $i$ is the player index, $j$ is the year index and $k$ is the
innings index so that $X_{ijk}$ represents the number of runs scored
by player $i$ in his $j$th year during his $k$th innings of that year.
Also $n_i$ and $n_{ij}$ denote respectively the number of years in the
career of player $i$ and the number of innings played during year $j$
in the career of player $i$.

Notation for other available information is as follows. For player~$i$
in the $k$th innings of their $j$th career year: $y_{ijk}$ is the year
the innings was played, $a_{ijk}$ is the age of the player, $h_{ijk}$
indicates whether the innings was played in the batsmen's home country
($1=\text{home}$, $2=\text{away}$), $m_{ijk}$ is a within-match
innings index (different from the within-year innings label $k$),
$o_{ijk}$ is the opposition country, and $e_{ijk}$ is an
  indicator for the era of play, which here is considered on a decade
  scale. These last two pieces of information together allow us to
  study possible changes to the performance of a country over time.

Within this Poisson framework we adopt a log-linear model for the run
scoring rate which includes the main components thought to
  influence its outcome, with
\begin{equation}\label{eq:loglinmodel}
\log\lambda_{ijk} = \theta_i + \delta_{y_{ijk}} + f_i (a_{ijk}) 
+ \zeta_{h_{ijk}} + \nu_{m_{ijk}} + \xi_{o_{ijk}} + \omega_{o_{ijk}, e_{ijk}}
\end{equation}
where $\theta_i$ represents the ability of player $i$, the difficulty
of the year is captured through $\delta_\ell$ (the data span \dataspan
years), and $f_i(a)$ is a player-specific ageing function, of which
more in a moment. The remaining terms in the model are game-specific,
representing respectively the effect of playing at or away from home,
the match innings effects, the quality of the opposition and an
interaction term allowing for the quality of the opposition to change
over different eras. Here we take eras to be decades to reduce the
number of parameters in the model.

The player ability parameter captures the contribution to runs scored
that can be attributed to the fundamental talent of the player. As
mentioned earlier, ageing can have a strong impact upon sporting
performance so we incorporate an individual quadratic ageing function as
suggested by Albert's discussion in~\cite{BRL}, namely
\begin{equation*}
\label{eq:agefunct}
f_i(a) = -\alpha_{2i}(a-\alpha_{1i})^2,
\end{equation*}
where $\alpha_{1i}$ is the age at which the peak is attained and
$\alpha_{2i}$ is the curvature of the function which measures the rate
at which the individual matures and declines.

The year effects are a composite of several factors: clearcut changes
such as depth of competition (more Test playing countries), game focus
(scoring rates are far higher in modern times and there are fewer
draws) and law changes (e.g fewer bouncers per over allowed to make
batting easier) whereas others are more subtle, for instance
technological advances and game conditions (most pitches are prepared
to last five days to ensure maximum profit). We anticipate that the
year effects vary smoothly over time and allow for this by using a
random walk prior; see Section~\ref{sec:prior}. The year effects also
need to be standardised for identifiability reasons and so we compare
these effects relative to the final year in the dataset
(\finaltestyear) by taking $\delta_{\dataspan}=0$.

The remaining terms in the model account for home advantage, which is
common in many team sports, and two further context-specific effects
to represent that as pitches deteriorate, and the match situation
becomes more acute, batting may become more difficult and to take into
account the quality of the opposition. We set the home effect as the
reference level (by taking $\zeta_1 =0$) and measure the impact of
playing away by $\zeta_2$. The innings effects are represented through
$\nu_g$ to reflect the difficulty of innings $g$ where $g = 1, 2, 3,
4$ is the innings of the match in which the runs were scored, and with
$\nu_1=0$ for identifiability. The quality of the opposition is
taken into account via $\xi_q$ for $q = 1, \ldots, 10$ to represent
the ten Test playing countries, some of which have traditionally been
stronger than others. Here we number the countries alphabetically.
For reasons of identifiability we will take Australia, the first team
in the alphabetical ordering of the Test match playing nations, as the
reference opposition country, with $\xi_1=0$. Further, the
  opposition-decade interactions are compared to the final decade (by
  taking $\omega_{1:10,14}=0$) and to that of Australia (by also taking
  $\omega_{1,1:13}=0$).

Thus in this model $\exp(\theta_i)$ is the average number of runs per
innings scored by player $i$ when he is at his peak age, playing at
home against Australia, and in the first innings of a Test match
taking place in \finaltestyear.

\subsection{Poisson random effects model}
There is substantial variation in individual innings-by-innings
cricket scores. As such, the inherent assumption of equidispersion in
the Poisson model is unlikely to hold. Under this model and
considering players that score on average ten or more runs per innings
one would expect their distribution of scores to be broadly Gaussian.
However any follower of cricket would intuitively feel that this is
not the case and that excess variability to that provided by the
Poisson model is present. The data in Fig.~\ref{fig:IanBell} on Ian
Bell are typical of many other players and show extra-Poisson
variation with censored observations and perhaps more ducks than
anticipated. We now augment the model to allow for each of these
features.

We allow for the extra-Poisson variation by introducing random
effects, acting multiplicatively on the Poisson mean parameter, so
that
\[
X_{ijk}|\lambda_{ijk},v_{ijk}\indep\Po(\lambda_{ijk} v_{ijk}).
\]
There are many possible choices of distribution for the random effects
$v_{ijk}$, such as gamma, log-normal, inverse Gaussian or general
power transforms \citep{Hougaard}. We will use the gamma distribution
as this gives a negative binomial distribution for the number of runs
after integrating over $v_{ijk}$ \citep{CameronTrivedi86, Greene2008}.
This choice allows a direct comparison with earlier work, particularly
as the geometric distribution is a special case. For further
flexibility we allow the random effects distributions to be
player-specific, reflecting that player characteristics, such as
aggression, can lead to substantial differences in variability between
players of comparable ability. Thus we take $v_{ijk}\sim\Ga(\eta_i,
\eta_i)$, with $E(v_{ijk})=1$ and $Var(v_{ijk})=1/\eta_i$. Therefore
(marginally) we use a negative binomial model for runs scored, with
\[
X_{ijk}|\lambda_{ijk},\eta_i \indep NB\{\eta_i,\eta_i/(\eta_i+\lambda_{ijk})\}.
\]
Note that introducing the random effects makes no change to the
(marginal) mean but has inflated the (marginal) variance to $
Var(X_{ijk})=\lambda_{ijk}(1+\lambda_{ijk}/\eta_i), $ with the basic
Poisson model being recovered as $\eta_i\rightarrow\infty$. This form
of variance function is appropriate for modelling batting scores as
the variability is smaller for players of lesser ability (they have a
more restricted range of runs scored and rarely achieve high numbers
of runs) and larger for players of higher ability (although they score
high numbers of runs, they will typically also have innings with very
low scores).  The form of the negative binomial success probability is
cumbersome and so to simplify the exposition we will use
$\beta_{ijk}=\lambda_{ijk}/\eta_i$.

We now augment the model to deal with (i) incomplete scores -- innings
where the batsmen is not dismissed; (ii) potential zero-inflation in
the data -- more ducks (zero scores) in the data than the model
suggests.

\subsubsection{Censoring}\label{Subsec:censoring}
Approximately $13\%$ of innings are incomplete, referred to as `not
out' in cricketing vernacular, and are typically due to the completion
of a team innings, which, by necessity, must include one incomplete
innings at the fall of the final wicket, or two incomplete innings in
a successful run chase (or if the match has not been completed due to
adverse weather or running out of time). Incomplete innings can also
happen when the team captain `declares' and brings the innings to a
premature close (typically to aid the prospect of victory) and this
can result in either one or two incomplete innings. Historically,
cricket has dealt with incomplete innings in a somewhat \textit{ad
  hoc} manner whereby the runs are added to the numerator in the
batting average without any increment to the denominator. Clearly such
innings ought not to be dealt with in the same way as a complete
innings and the standard cricketing treatment can exaggerate the
contribution of incomplete innings and thereby affect the batting
average. From a statistical viewpoint, a not-out is simply a censored
observation. \cite{Kimber} claim that `$x$ not-out is representative
of all scores of $x$ or more' and so we assume non-informative
censoring.
Thus, denoting a not-out (censored) innings by the binary
variable $c$, the likelihood contribution from player $i$, for the
$k$th innings of the $j$th year of his career, is
\begin{equation}\label{Eq:CensLike}
\left\{\binom{x_{ijk}+\eta_i-1}{x_{ijk}}
\beta_{ijk}^{x_{ijk}}/(1+\beta_{ijk})^{\eta_i+x_{ijk}}\right\}^{1-c_{ijk}} 
\times P(X_{ijk}\geq x_{ijk})^{c_{ijk}},
\end{equation}
where $X_{ijk}$ has a negative binomial distribution.

\subsubsection{Zero-inflation}
After ignoring censored zeroes, ducks account for almost $11\%$ of the
observations in the data.  Even Sir Donald Bradman (with a Test
batting average of 99.94) had a modal score of zero with seven ducks
out of eighty innings.  This high proportion of zeroes is likely to be
due to players being vulnerable early in their innings \citep{Brewer},
taking time to acclimatise to conditions and `get their eye in' rather
than owing to some other process that causes scores to be necessarily
zero. Thus the proportion of ducks is likely to be higher than
expected using the Poisson random effects model and so we modify the
model to allow for this inflation of zeros.  We also allow for
different levels of zero-inflation for each player.

There are two basic ways of dealing with zero inflation. One way is to
model the probability of getting a zero by a mixture of the primary
model and a point mass at zero \citep{Lambert} and the other is to use
a hurdle model which contains a model for zero counts (the hurdle
component) and a separate model for the strictly positive counts (once
the hurdle, a batsmen playing a scoring stroke for instance, has been
cleared). Hurdle models are particularly popular in the economics
literature; see, for example, \cite{Gurmu97, Gurmu98}. They are the
natural choice when the zeroes are entirely structural, such as in a
biological process \citep{Ridout} or a weather pattern \citep{Scheel}.
We favour the mixture representation as this can be interpreted as the
number of ducks being a mixture of (the Poisson random effects) model
based zeros and a component representing the increased vulnerability
of a batsman early in an innings. This representation has the
additional advantage (not followed up here) of providing a framework
for generalising the model to inflate other scores, such as four or
six, that may occur more frequently due to being achievable with a
single scoring stroke, that is, via a `four' or a `six'.

The excess zeroes are assumed to be unrelated to the other effects and
so we model the probability of getting a (completed) duck for player $i$ as
\[
Pr(X_{ijk} = 0) = \pi_i + (1 - \pi_i)/(1+\beta_{ijk})^{\eta_i}.
\]
Note that as the player-specific parameter $\pi_i\rightarrow 0$, the
zero-inflated component diminishes and the number of (completed) ducks
is well described by an orthodox Poisson random effects model. Thus,
denoting a batsman with a (completed) duck by the binary variable $d$,
the likelihood contribution from player $i$, for the $k$th innings of
the $j$th year of his career, is amended from that in
(\ref{Eq:CensLike}) to
\begin{equation}
\begin{split}
&\left\{\pi_i+(1-\pi_i)/(1+\beta_{ijk})^{\eta_i}\right\}^{(1-c_{ijk})d_{ijk}}\\
&\qquad\times\left[(1-\pi_i)\left\{\binom{x_{ijk}+\eta_i-1}{x_{ijk}}
\beta_{ijk}^{x_{ijk}}/(1+\beta_{ijk})^{\eta_i+x_{ijk}}\right\}^{1-c_{ijk}} 
P(X\geq x_{ijk})^{c_{ijk}}\right]^{1-d_{ijk}}.
\end{split}
\label{Eq:CensZIPLike}
\end{equation}
Introducing a zero-inflation effect also reduces the expected number
of runs scored by a factor of $(1-\pi_i)$.

\section{The prior distribution}\label{sec:prior}
We need to construct a joint prior distribution for the many
parameters in this model. In general, we have chosen to describe our
prior beliefs by taking fairly weak independent priors for each
parameter component. This has the benefit of ``letting the data
speak'' and gives our results a reasonable level of robustness against
our choice of prior.

We adopt a random effects style (or hierarchical) prior for the
player-specific ability parameters in which ability varies between
batsmen by taking
\[
\theta_i\mid\mu_\theta,\sigma_\theta
\indep N\bigl(\mu_\theta,\sigma_\theta^2\bigr).
\]
We also take semi-conjugate prior distributions for the ability
parameters, with $\mu_\theta\sim N(m_\mu,s_\mu^2)$ and
$\sigma^2_\theta\sim IG(a_\sigma,b_\sigma)$, where $IG(a,b)$ denotes
the Inverse Gamma distribution with mean $b/(a-1)$.  It was felt that
the median number of runs scored across all innings (including
not-outs) would be around 20 and so we take $m_\mu=\log 20$. Also the
variability between decades of runs scored was likely to be within a
60\%-fold increase or decrease and so we take $s_\mu=0.25$ (as
$e^{0.5}\simeq 1.6$). Variation of player ability was thought to be
typically about a four fold increase/decrease around the decade mean,
giving $\sigma^2_\theta$ a mean of around $0.5$, and that the
probability that this fold increase/decrease would exceed ten was
around 5\%.  Together these requirements give a prior distribution
with (roughly) $a_\sigma=3$ and $b_\sigma=1$.

It was felt that the year effects $\delta_\ell$ should vary fairly
smoothly in time and that prior beliefs were less certain for years
going further and further into the past. Therefore, together with the
identifiability constraint $\delta_{\dataspan}=0$, we use the
(backwards) simple random walk
$\delta_\ell=\delta_{\ell+1}+\sigma_\delta\epsilon_\ell$,
$\ell=\numnonzerodelta,\ldots,1$, where the $\epsilon_\ell$ are
independent standard normal quantities. To see its smoothing role, it
is useful to note that this random walk induces
\begin{equation*}\label{Eq:YearPrior}
\delta_\ell \mid \deltavec_{(\ell)}, \sigma_\delta \sim N\left(\frac{\delta_{\ell - 1} + \delta_{\ell + 1}}{2},  \frac{\sigma_\delta^2}{2} \right)
,\qquad\text{for }\ell=2,\ldots, \numnonzerodelta,
\end{equation*}
with $\delta_1\mid\deltavec_{(1)},\sigma_\delta\sim
N(\delta_2,\sigma_\delta^2)$, where
$\deltavec_{(\ell)}=(\delta_i,~i\neq \ell)$ represents all of the year
effects except year~$\ell$. For notational convenience we write
$\deltavec$ for the year effects $\deltavec_{(\dataspan)}$.  These
descriptions lead to the prior distribution of the year effects being
$\deltavec\mid\sigma_\delta\sim N_{\numnonzerodelta}(\mathbf{0},
\sigma_\delta^2\,Q^{-1})$ where the inverse correlation matrix $Q$ has
the tri-diagonal structure
\[
Q=\begin{pmatrix}
1& -1& & &\\
-1 & 2 & -1 &\\
&\ddots& \ddots& \ddots&  \\
&& -1& 2& -1\\
&& & -1& 2
\end{pmatrix}.
\]

The parameter $\sigma_\delta$ describes the smoothness of the year
effects and, as this impacts player ability on an exponential scale,
it was felt that $\sigma_\delta^2$ should have an
$IG(a_\delta,b_\delta)$ prior distribution with mean 0.01 and only a
5\% probability of exceeding 0.03. This leads (roughly) to a choice of
prior parameters $a_\delta=2$ and $b_\delta=0.01$.

We now consider the prior distributions for the remaining parameters,
beginning with the game-specific parameters: the effect of playing
away $\zeta_2$, the innings effects $\nu_{2:4}$, the quality of the
opposition $\xi_{2:10}$ and the opposition-era interactions
$\omega_{2:10,1:13}$ (recall that
$\zeta_1=\nu_1=\xi_1=\omega_{1:10,14}=\omega_{1,1:13}=0$ for
identifiability). The strength of our opinion on their potential size
is quite weak and so we give these parameters zero mean normal prior
distributions with standard deviation 0.5, this taken to equate to a
95\% prior credible interval for these effects spanning an
increase/decrease of around 2.7--fold on the runs scored.
Our prior beliefs about the player-specific ageing function are that
the peak age is around 30 years old and that the rate of
maturity/decline of players at seven years before (after) their peak
is roughly 2/3 ($-2/3$). We represent our fairly weak prior beliefs by
taking $\alpha_{1i}\sim N(30,4)$ and $\alpha_{2i}\sim LN(-3,9)$.

Previous studies have considered a geometric random effects
distribution for runs scored and so we give the individual random
effects heterogeneity parameters $\eta_i$ a log-normal prior with unit
prior median, but also make this prior fairly weak by taking
$\eta_i\sim\LN(0,1)$. Our prior beliefs about the individual
zero-inflation parameters $\pi_i$ are captured by a
$Beta(a_\pi,b_\pi)$ distribution with mean 0.1 and only a 5\%
probability of $\pi_i$ exceeding 0.3. This leads (roughly) to a choice
of prior parameters $a_\pi=1$ and $b_\pi=9$.

\section{The posterior distribution}
\label{sec:posterior}
The posterior density can be factorised as
\[
\pi(\kappavec,\etavec,\pivec\mid\xvec,\cvec,\dvec) \propto 
\pi(\xvec,\cvec,\dvec\mid\kappavec, \etavec, \pivec) \pi(\kappavec)\pi(\etavec) \pi(\pivec) 
\]
with $\lambdavec=\lambdavec(\kappavec)$, where $\xvec$, $\cvec$ and
$\dvec$ are the vectors of runs scored and associated censoring and
duck indicators respectively, and
$\kappavec=(\thetavec,\deltavec,\sigma_\delta,\alphavec,
\zeta_2,\nuvec,\xivec,\omegavec)$ contains the remaining parameters in
the model, with $\nuvec=(\nu_{2:4})$, $\xivec=(\xi_{2:10})$, and
$\omegavec=(\omega_{2:10,1:13})$. This posterior distribution is
analytically intractable and we therefore turn to a sampling-based
approach and make inferences via the use of Markov chain Monte Carlo
(MCMC) methods.

In our MCMC scheme we generally use Metropolis-Hastings steps with
symmetric normal random walk proposals on an appropriate scale and
centred on the current value; for example, on the log scale for
positive quantities or the logit scale for quantities restricted to
$(0,1)$. Overall we have found this strategy to work well except for
updates to the year effects~$\deltavec$. Here Gibbs updates are
available for each component $\delta_\ell$ but their full conditional
distributions depend strongly on the values taken by the year effects
on either side, that is,
$\pi(\delta_1|\cdot)=\pi(\delta_1|\delta_2,\cdot)$ and
$\pi(\delta_\ell|\cdot)=\pi(\delta_\ell|\delta_{\ell-1},\delta_{\ell+1},\cdot)$,
$\ell\neq 1$. This is not surprising given the dependence structure in
the prior distribution for~$\deltavec$. It is well known that such
strong dependence can lead to poor mixing such as that in, for
example, the distribution of hidden states in hidden Markov models.
Also this strong dependence prohibits using software such as JAGS
\citep{Plummer2004} to obtain posterior realisations in a timely
manner. Instead we follow \citet{Gamerman1997} and construct a normal
proposal distribution for $\deltavec$ via a Taylor series
approximation to its full conditional distribution; see the
supplementary materials for further details. We have found this
strategy to greatly improve the mixing of the scheme.

\section{Results}\label{sec:results}
An implementation of the MCMC scheme in R~\citep{Rlang} is available
from\\ \verb$https://github.com/petephilipson/Ranking-Test-batsmen$ together with
the data. We report here results from a typical run of the MCMC scheme
which used a burn-in of 5K iterations and was then run for a further
200K iterations, with this (converged) output thinned by taking every
$20$th iterate. This gave a posterior sample of $N=10$K (almost)
un-autocorrelated values for analysis. Convergence was assessed
through a variety of graphical and numerical diagnostics via the R
package coda~\citep{Rcoda}.

\subsection{Random effects distribution for player ability}
\begin{figure}[t!]
\centering
\includegraphics[width=0.45\textwidth]{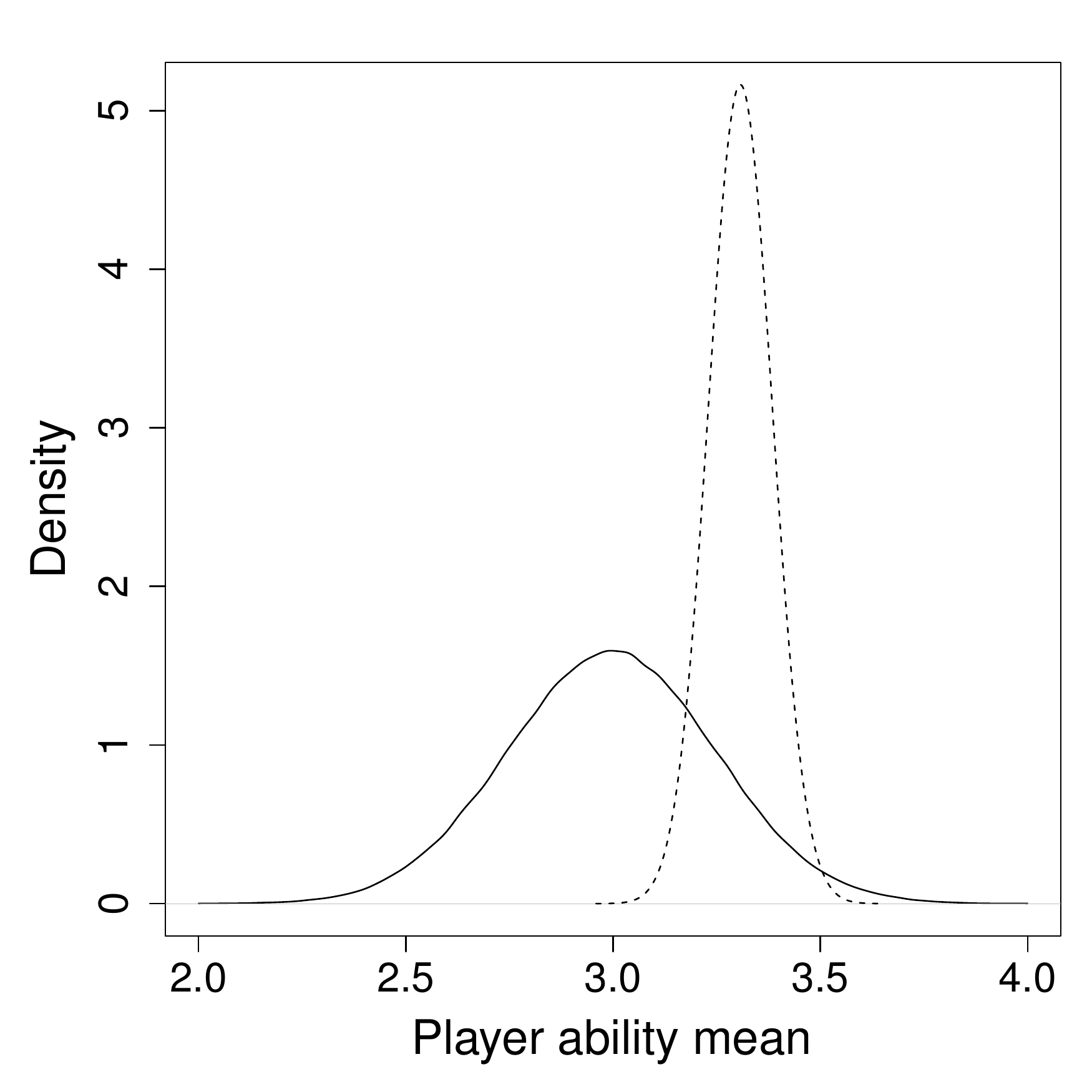}
\includegraphics[width=0.45\textwidth]{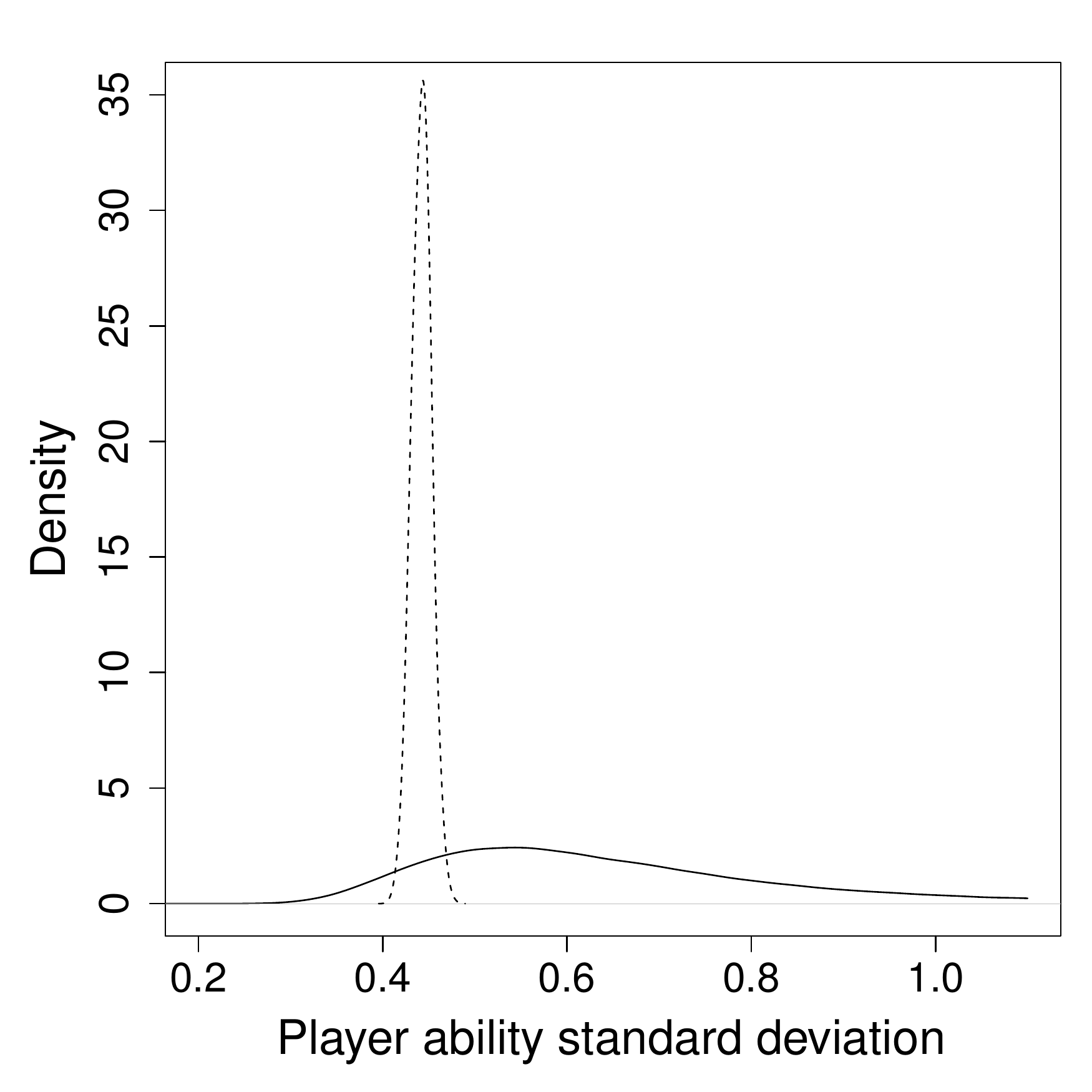}
\caption{Prior (solid line) and posterior (dashed line) density plots
  for ability parameters $\mu_\theta$ (left plot) and $\sigma_\theta$
  (right plot).}
\label{FigHyper}
\end{figure}

The (marginal) posterior distributions for the mean and standard
deviation ($\mu_\theta$ and $\sigma_\theta$) of the random effects
distribution for player ability are shown in Fig.~\ref{FigHyper}.
Clearly the data have been quite informative. We can get a quick
understanding of this posterior distribution by looking at its
implication for the (random effects) distribution of the number of
runs scored (by players at their peak age, playing at home against
Australia, and in the first innings of a Test match taking place in
\finaltestyear). Ignoring the (player-specific) zero-inflation effect,
the five number summary (Min--LQ--Med--UQ--Max) for the median number
of runs scored ($\exp(\mu_\theta)$) is $24.7-26.4-27.3-28.3-30.2$, and
that for the average number of runs scored
($\exp(\mu_\theta+\sigma_\theta^2/2)$) is $25.2-26.9-27.9-28.8-30.8$.
These distributions seem reasonable after taking into account that the
zero-inflation parameters $\pi$ are around 8\% (see
section~\ref{Subsec:ModelExtensionResults}).

\subsection{Year effects}
\begin{figure}[t]
\centering
\includegraphics[width=0.65\textwidth]{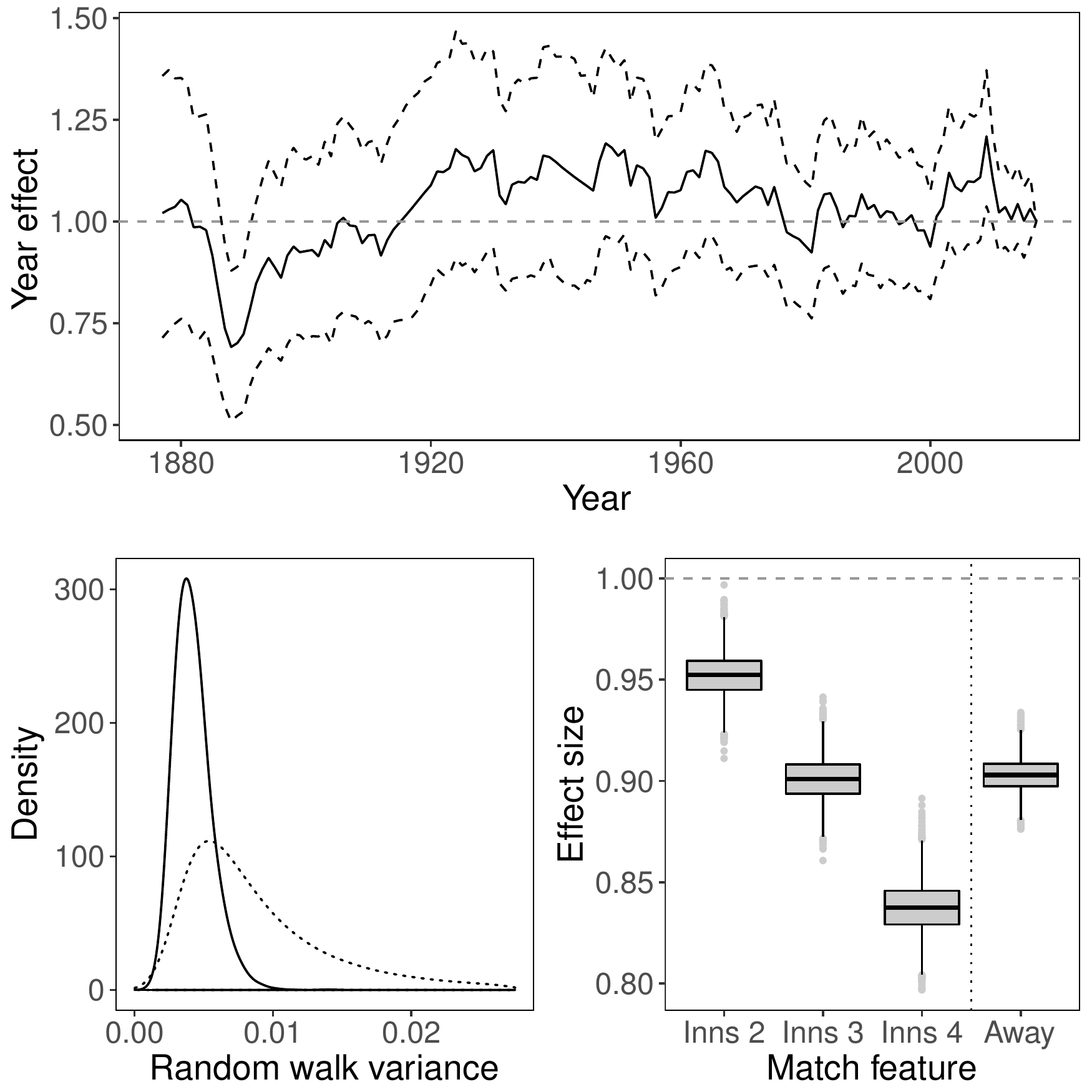}
\caption{Top: posterior mean and central 95\% bands for the
  multiplicative year effects exp($\delta_\ell$). Bottom left:
  posterior (solid) and prior (dashed) density plots for the smoothing
  parameter $\sigma_\delta^2$. Bottom right: boxplots of the posterior
  distributions for the multiplicative match innings effects
  exp($\nu_m$) and playing away effect exp($\zeta_2$).}
\label{FigYear_RWprior_InnsHA}
\end{figure}

The posterior distribution for the year effects is summarised in the
upper panel of Fig.~\ref{FigYear_RWprior_InnsHA}.
The effects are shown on an exponential scale and so represent the
multiplicative effect on run scoring for each year, relative to
playing against Australia in the most recent year (\finaltestyear),
here shown by the horizontal dashed line. It is clear that there are
very few important year effects, with the main (and negative)
deviation being 1887--1891, a period when it is widely acknowledged
that bowling conditions were favourable. The next strongest (and
positive) deviation occurred in 2009, a year featuring four of the
sixteen highest team scores of all time. The bottom-left panel of
Fig.~\ref{FigYear_RWprior_InnsHA} shows the prior and posterior
distribution of the smoothing parameter $\sigma_\delta^2$ for the year
effects. The slight shift in the posterior towards lower values
suggests that the prior distribution did not over smooth.

\subsection{Home advantage, innings and opposition effects}
The bottom-right panel of Fig.~\ref{FigYear_RWprior_InnsHA}
  provides a visual comparison of the size of the multiplicative
  effect on run-scoring when batting in different innings and playing
  away from home. Note that these effects are all relative to playing
  at home against Australia in the first innings in \finaltestyear,
  represented by the dashed horizontal line.  The effect of playing
  away from home on runs scored, $\exp(\zeta_2)$, has posterior mean
  $0.90$ and 95\% confidence interval $(0.89, 0.92)$.  Thus, there is
  a pronounced detrimental effect of playing away from home, leading
  to batsmen scoring 10\% fewer runs. This finding is consistent with
  that found for `home advantage' in other sports \citep{Pollard2005,
    Jones2007, Baio2010}.  The posterior means (with 95\% confidence
  intervals) for the second, third and fourth innings effects
  ($\exp(\nu_{2:4})$) are $0.95\;(0.93, 0.97)$, $0.90\; (0.88, 0.92)$
  and $0.84\; (0.81, 0.86)$ respectively, with the reference value of
  one for the first innings.  These effects act multiplicatively on
  run-scoring. Hence, performance decreases as the match goes on, with
  the innings effect at its strongest in the final innings of the
  match, as cricketing folklore would have predicted. The second
  innings of a Test match is tougher than the first innings with a
  reduction of $5\%$ in runs scored, but the effect increases to a
  $10\%$ reduction in runs scored in the third innings and a $16\%$
  reduction in the final innings (compared to the first innings). It's
  interesting to see that the effect of batting in the third innings
  and that of playing away from home are very similar.

\begin{figure}[t]
\centering
\includegraphics[width=0.6\textwidth]{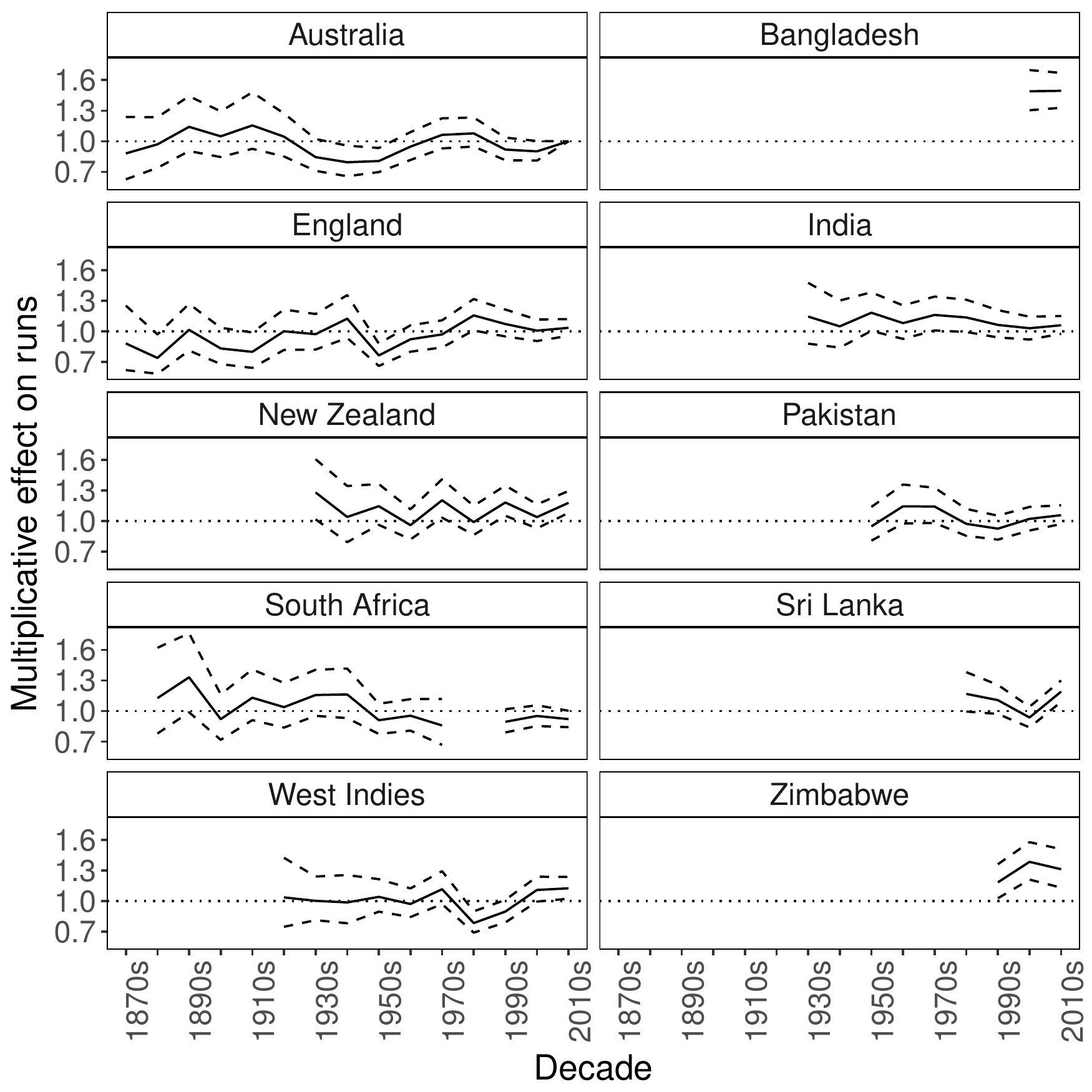}
\caption{Posterior mean and central 95\% bands for the multiplicative
  opposition effects by decade, exp($\xi_q + \omega_{qd}$).}
\label{FigDynamicOpp}
\end{figure}

Fig.~\ref{FigDynamicOpp} displays the posterior means and associated
95\% intervals of $\exp(\xi_q+\omega_{qd})$ for the ten Test playing
countries ($q=1,\ldots,10$) over the fifteen decades ($d=1,\ldots,15$)
during which Test cricket has been played. As mentioned earlier, fewer
countries played Test cricket when it commenced as an international
sport. The estimates in each case are relative to the strength of the
current Australian Test team (represented by the horizontal dotted
line on each plot). There are twenty instances of opposition effects
that show appreciable deviation from that of Australia in the most
recent decade: these are split as six instances of an opposition being
significantly more difficult to score runs against than the current
Australia team and fourteen cases where the opposition are easier to
score runs against.  The largest deviations (and with the lowest
posterior means) were England in the 1880s and 1950s, and the West
Indies in the 1980s, each causing a 20-25\% reduction in average runs
scored.

The two newest Test playing nations, Bangladesh and Zimbabwe, have
struggled at times to be competitive and the three largest
(significant) posterior means are for these two countries.  Batsmen
have preyed on this weakness, scoring on average over 50\% more runs
against Bangladesh and over 30\% more runs against Zimbabwe.  New
Zealand were also relatively weak when they first played Test cricket
(in the 1930s) with batsmen scoring on average 30\% more runs.  Other
noteworthy examples of weaker opposition were India in the 1950s,
India and New Zealand in the 1970s and England in the 1980s. In each
case batsmen scored on average around 20\% more runs against these
countries in these decades.

We investigated the sensitivity of our conclusions on opposition
  effects to using five year time periods rather than decades and
  found very little difference. Also there is no need to standardise
opposition scores against the current Australia side and it is
straightforward to standardise scores against any opposition team in
any decade. We provide the results for any choice
of team and decade via an RShiny application, available at\\
\verb$https://petephilipson.shinyapps.io/opposition/$

\subsection{Random effects heterogeneity and zero-inflation}
\label{Subsec:ModelExtensionResults}
Five number summaries of the posterior means and standard deviations
for the player-specific random effects heterogeneity parameters
($\eta_i$) are $0.48-0.87-1.01-1.16-2.19$ and
$0.07-0.24-0.40-0.54-0.89$ respectively. The posterior distributions
for a number of batsmen show clear deviation from the geometric model
$(\eta_i=1)$ for cricket scores postulated by \cite{Elderton1945} and
\cite{Wood1945}. We note that these authors did not account for
zero-inflation (or censoring) but Wood did remark on a lack of fit for
scores of zero.

Five number summaries of the posterior means and standard deviations
for the player-specific zero-inflation parameters ($\pi_i$) are
$0.01-0.06-0.08-0.11-0.34$ and $0.01-0.04-0.06-0.08-0.15$
respectively. The posterior distributions show clearly both evidence
for zero inflation in Test match cricket scores and variation between
players. The modal batsman's score in Test cricket is zero, and the
commonly held belief that batsmen are at their most vulnerable at the
onset of their innings is a plausible explanation here.  Posterior
means of the $\pi_i$ for the top thirty ranked batsmen are included in
Table~\ref{TableOfRanks}.  The excess of zeroes observed by Wood is a
genuine feature of Test cricket scores. It is interesting to note the
discussion on the use of the standard cricket batting average summary
in \cite{Kimber}: they point out that such a measure is only a
consistent estimate if the scores follow a geometric distribution.

\subsection{Individual ageing}
\begin{figure}[t]
\centering
\includegraphics[width=0.7\textwidth]{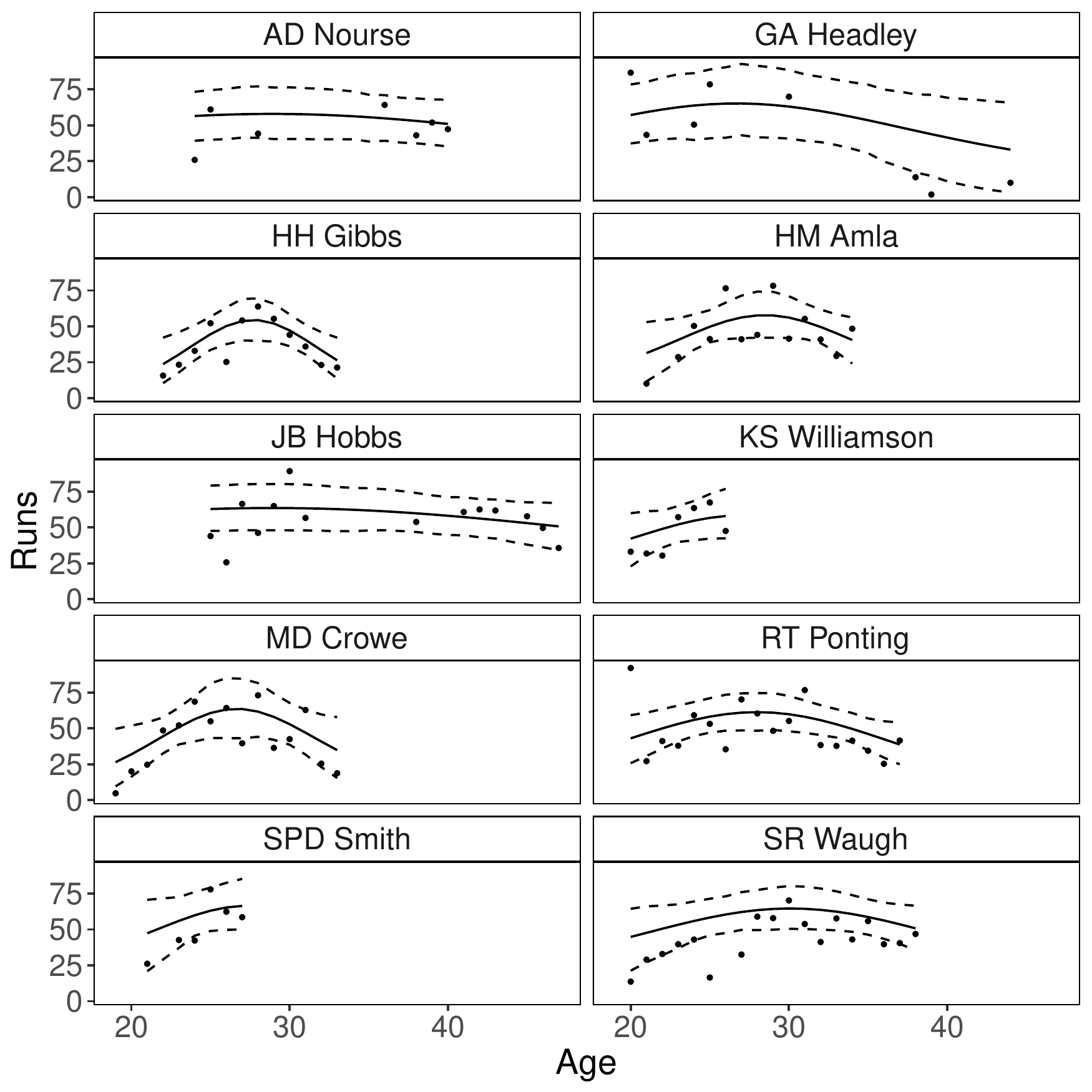}
\caption{Posterior mean and central 95\% bands for the ageing profile
  $(1-\pi_i)\exp\{\theta_i+f_i(a)\}$ of various batsmen together with
  their posterior mean adjusted runs scored.}
\label{PlayerProfiles}
\end{figure}

We determine the ageing profile for a batsman by examining the
posterior distribution of their expected runs scored at various
ages~$a$, that is, $(1-\pi)\exp\{\theta+f(a)\}$.
Fig.~\ref{PlayerProfiles} shows posterior mean profiles (and central
95\% bands) for a selection of players of broadly similar ability but
with quite different ageing profiles. Also included in the plot are
the posterior mean adjusted runs scored for each player, that is, the
posterior mean of
\[
\sum_{j,k:a_{ijk}=a} x_{ijk} \times \exp\{-(\delta_{y_{ijk}} + \zeta_{h_{ijk}} + \nu_{m_{ijk}} + \xi_{o_{ijk}} + \omega_{o_{ijk}, e_{ijk}})\}/n_{ia}
\]
where $n_{ia}$ is the number of completed innings played by player $i$
at age $a$. The figure shows that the quadratic function largely
captures the ageing profiles, particularly when taking account of the
(posterior) uncertainty on the mean adjusted scores (not shown). The
posterior mean of the peak ages ($\alpha_{1i}$) is typically late
twenties; see Table~\ref{TableOfRanks}.

\subsection{Player rankings}
\begin{figure}[t]
    \centering
    \includegraphics[width=0.75\textwidth]{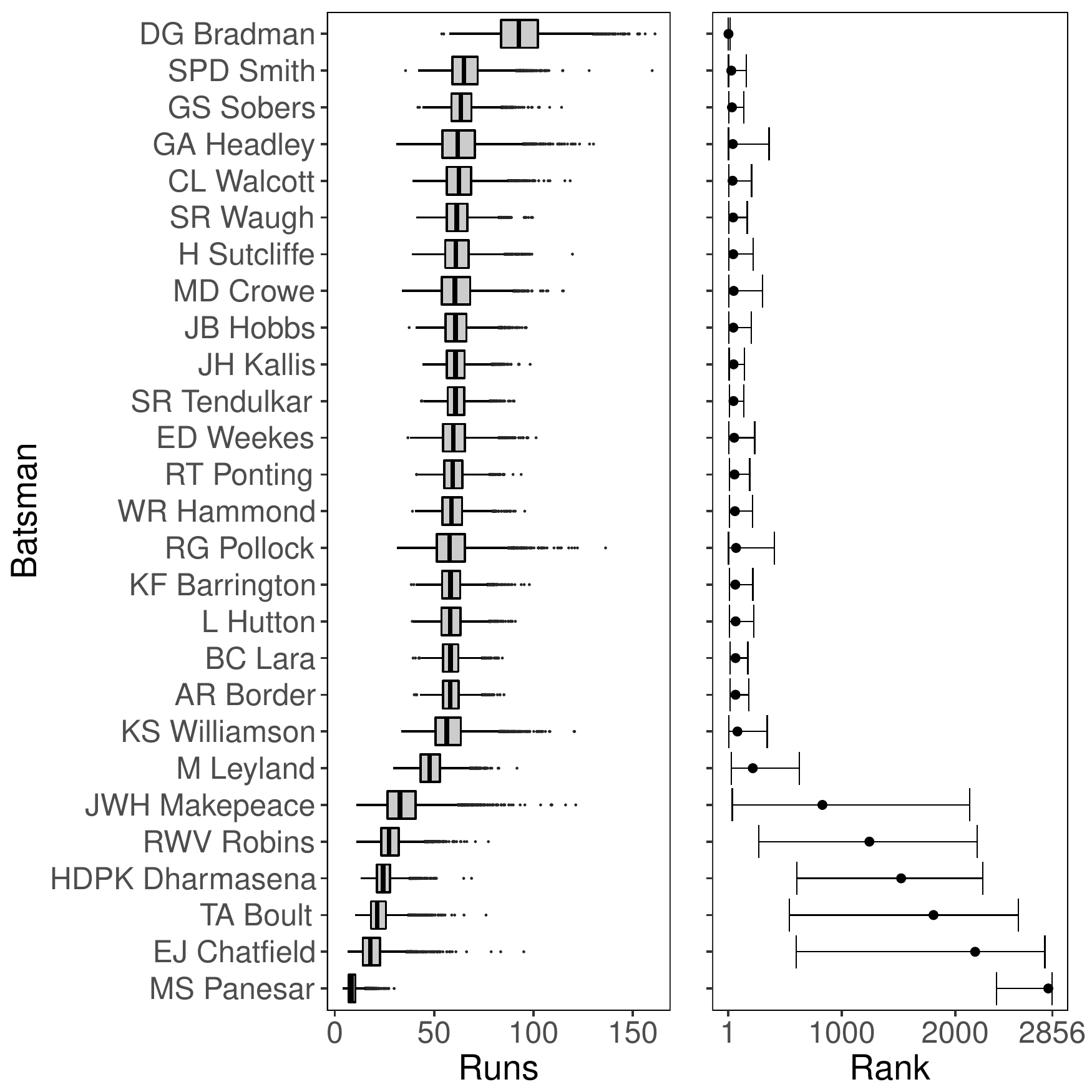}
    \caption{Posterior distributions of some player abilities
      exp($\theta_i$) and their ranking confidence intervals.}
\label{FigRunsRanks}
\end{figure}

The posterior distribution of mean runs scored by the top thirty
ranked players are shown as boxplots in Fig.~\ref{FigRunsRanks}, with
numerical summaries in Table~\ref{TableOfRanks}.  Here the players are
listed by their posterior mean of $(1-\pi)\exp(\theta)$, that is,
their expected runs scored at their peak age assuming the year of play
is \finaltestyear (no year effect) and batting at home in the first
innings of a Test match against Australia. It is striking just how far
Sir Donald Bradman is ahead of the other batsmen, in terms of
posterior mass; his extraordinary average is well-known to cricket
fans and the plot captures this clearly.  The posterior distributions
of the players ranked from two to thirty are largely similar, with
considerable overlap. After Bradman it is unclear who is the next
`best' batsman. This point is further underlined by the posterior
distribution of each player's rank, calculated across the MCMC
samples. The figure also shows the median posterior rank, together
with equi-tailed 95\% confidence intervals. The numerical summaries
for each batsman can be found in Table~\ref{TableOfRanks}.  Note that
these are summaries of marginal distributions for each player and not,
for example, the most probable joint ranking across all players.
Therefore it is possible, and happens here, that no batsman has a
posterior median rank of, say, two. However, given the level of
variation in runs scored, it does not seem reasonable to rank batsman
by a single number summary, be it mean score or median rank.
\cite{Kimber} make a similar argument, stating that `it is clear that
a one-number summary of the distribution of a batsman's scores is not
enough'. Our rank confidence intervals give a much more reasonable
measure of rank position and its uncertainty.

The interval for Bradman's rank is quite narrow, ranging from rank~1
to rank~14.  There is little difference in the career batting averages
of many players after Bradman and this is borne out in the spread of
the confidence intervals for the rankings, which are largely similar
and noticeably wide. It is interesting to see the level of (posterior)
uncertainty on player rankings.  Fig.~\ref{FigRunsRanks} shows
confidence intervals for the top twenty players along with players
ranked $100$th, $500$th and every five-hundredth player thereafter up
to the $2500$th player and the final player, ranked
$\numtestcricketers$th. The high level of posterior uncertainty in
these ranks chimes with a remark by \cite{Goldstein96} when comparing
institutional performances, that `such variability in rankings appears
to be an inevitable consequence of attempting to rank individuals with
broadly similar performances'.  A full list of the ability scores
  and ranks for all \numtestcricketers batsmen can be found via the
  RShiny application at
\verb$https://petephilipson.shinyapps.io/BatsmenRankings/$

\begin{table}
  \caption{\label{TableOfRanks}Player rankings (ordered by posterior
    mean runs at peak age) together with posterior means for peak age and zero-inflation
    proportions, and summaries of player rank distributions.}
\centering
\fbox{%
\begin{tabular}{c l r c r r c  c l}
& & & & & & Peak & Zero- & Median rank\\
Rank& Name&  Debut& Innings& Runs& SD& age& inflation& (95\% CI) \\
\hline
\d 1& DG Bradman& 1928& \d 80& 93.7& 14.3& 28.2& 7\%& \d 2 (1-14)\\ 
\d 2& SPD Smith& 2010& 100& 66.3& 10.1& 27.9& 2\%& 27 (3-158)\\ 
\d 3& GS Sobers& 1954& 160& 64.1& 7.9& 27.8& 5\%& 33 (5-137)\\ 
\d 4& GA Headley& 1930& \d 40& 63.2& 12.8& 27.0& 4\%& 40 (2-360)\\ 
\d 5& CL Walcott& 1948& \d 74& 63.2& 9.6& 28.4& 2\%& 38 (4-206)\\ 
\d 6& SR Waugh& 1985& 260& 62.0& 7.7& 29.6& 5\%& 43 (7-168)\\ 
\d 7& H Sutcliffe& 1924& \d 84& 61.9& 9.0& 28.2& 3\%& 44 (5-218)\\ 
\d 8& MD Crowe& 1982& 131& 61.6& 10.7& 27.0& 4\%& 48 (4-304)\\ 
\d 9& JB Hobbs& 1908& 102& 61.4& 8.2& 28.4& 4\%& 45 (6-204)\\ 
10& JH Kallis& 1995& 280& 61.3& 6.8& 28.6& 3\%& 47 (8-144)\\ 
11& SR Tendulkar& 1989& 329& 61.2& 6.4& 27.2& 2\%& 46 (10-139)\\ 
12& ED Weekes& 1948& \d 81& 60.5& 8.6& 27.8& 5\%& 52 (6-233)\\ 
13& RT Ponting& 1995& 285& 59.8& 6.8& 27.8& 3\%& 55 (10-189)\\ 
14& WR Hammond& 1927& 140& 59.4& 7.4& 28.1& 2\%& 59 (9-212)\\ 
15& RG Pollock& 1963& \d 41& 59.2& 11.7& 27.7& 3\%& 69 (3-406)\\ 
16& KF Barrington& 1955& 131& 58.9& 7.1& 28.6& 2\%& 62 (11-217)\\ 
17& L Hutton& 1937& 138& 58.7& 7.3& 28.1& 2\%& 65 (10-224)\\ 
18& BC Lara& 1990& 232& 58.6& 5.8& 28.3& 3\%& 64 (16-173)\\ 
19& AR Border& 1978& 265& 58.5& 6.0& 27.9& 2\%& 64 (14-180)\\ 
20& KS Williamson& 2010& 110& 58.0& 10.4& 27.7& 3\%& 81 (4-343)\\ 
21& Y Khan& 2000& 213& 57.7& 6.6& 28.7& 5\%& 73 (13-219)\\ 
22& KC Sangakkara& 2000& 233& 57.5& 5.8& 28.9& 2\%& 72.5 (16-198)\\ 
23& R Dravid& 1996& 286& 57.5& 5.7& 27.8& 1\%& 73 (17-198)\\ 
24& GS Chappell& 1970& 151& 57.4& 6.7& 28.2& 5\%& 75 (12-242)\\ 
25& AC Voges& 2015& \d 31& 57.2& 14.2& 28.5& 4\%& 91 (3-644)\\ 
26& HM Amla& 2004& 183& 57.2& 8.5& 28.5& 2\%& 80 (9-329)\\ 
27& JE Root& 2012& 107& 57.1& 7.8& 27.7& 2\%& 79 (9-303)\\ 
28& A Flower& 1992& 112& 56.9& 7.3& 28.5& 2\%& 80 (12-274)\\ 
29& SM Gavaskar& 1971& 214& 56.7& 6.1& 28.0& 2\%& 81 (18-226)\\ 
30& M Yousuf& 1998& 156& 56.6& 7.2& 28.6& 5\%& 87 (11-267)\\
\end{tabular}}
\end{table}

There are two established rankings lists with which we can compare our
rankings. The first is the traditional rankings by career Test batting
average and the second is the `Reliance ICC Best-Ever Test
Championship Rating' (ICC) list. These two differ in that the former
is a single measure across a player's entire career whereas the latter
is the maximum of a dynamic index which
places a greater emphasis on recent innings. Our approach could be
considered to be a compromise between these two systems. Overall, of
the top~30 in our rankings by posterior mean runs scored, we have 23
in common with all-time highest career batting average rankings, and
19 in common with the ICC rankings. Five batsmen in
Table~\ref{TableOfRanks} do not appear in either of these established
ranking lists; these batsmen (with our ranking by posterior mean runs
and 95\% confidence interval for their rank) are Waugh 6 $(7,168)$,
Crowe 8 $(4, 304)$, Border 19 $(14,180)$, Williamson 20 $(4,343)$ and
Flower 28 $(12,274)$. This illustrates a central problem in ranking
batsmen by a single number summary when there is a high level of
innings-to-innings variation in runs scored by each batsman.  In
particular the traditional ranking does not adjust for any covariate
information. The ICC rank does adjust for opposition/pitch effects but
is empirical and has some other \textit{ad hoc} adjustments.  Neither
system adjusts for the censoring (not-out) problem in a way that takes
account of player ability.

\subsection{Model fitting}
\begin{figure}[t]
\centering
\includegraphics[width=0.45\textwidth]{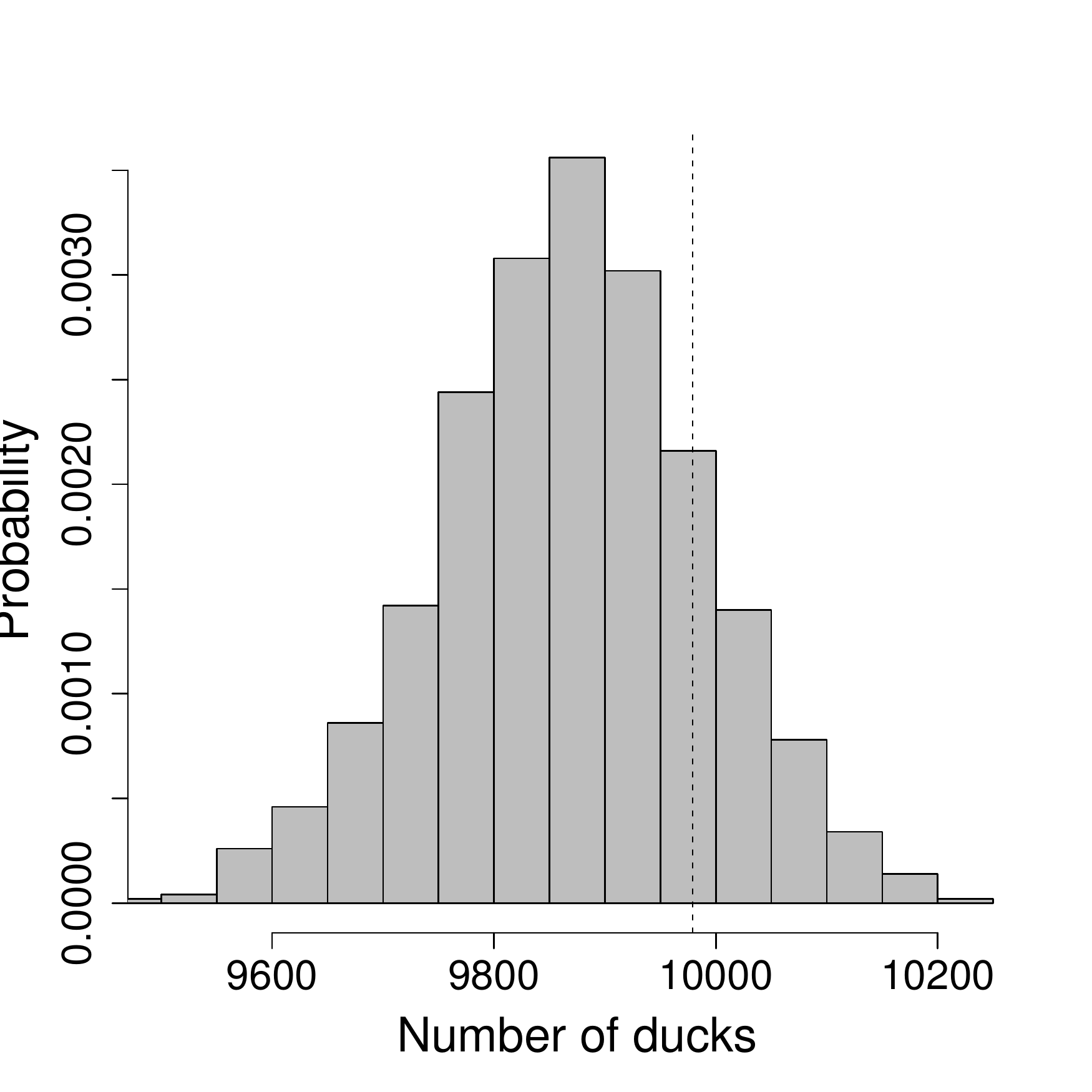}
\includegraphics[width=0.45\textwidth]{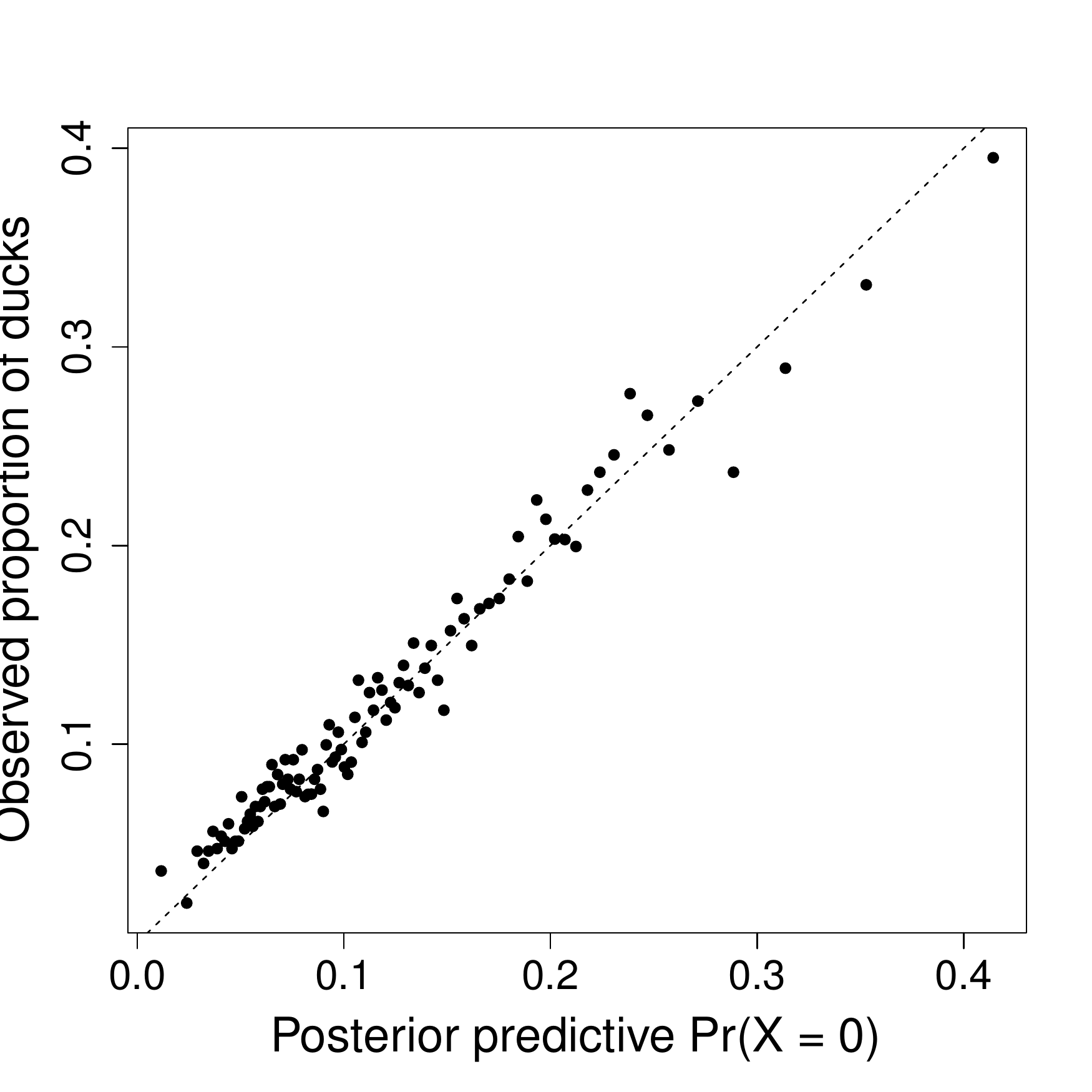}
\caption{Left: Posterior predictive distribution of total ducks (dashed line -- observed
  number in the data); Right: Observed proportion of ducks against centiles of
  posterior predictive probabiities of a duck.}
\label{FigOvE}
\end{figure}

We can study the ability of the model to predict ducks (zero scores)
by looking at the (model-based) posterior predictive probability of a
duck and seeing how this correlates with observed ducks. This
predictive probability is calculated by averaging a typical
model-based probability $Pr(X_{ijk}=0|\kappavec,\eta_i,\pi_i)$ over
the uncertainty in the posterior distribution. Therefore we estimate
these predictive probabilities using
\[
Pr(X_{ijk}=0|\xvec,\cvec,\dvec)=\frac{1}{N}\sum_{\ell=1}^N
Pr(X_{ijk}=0|\kappavec^{(\ell)},\eta_i^{(\ell)},\pi_i^{(\ell)}),
\] 
where
$\{\kappavec^{(\ell)},\etavec^{(\ell)},\pivec^{(\ell)}:\ell=1,\ldots,N\}$
is the posterior sample. Fig.~\ref{FigOvE} shows a summary of this
information. The left hand plot shows the (posterior) predictive
distribution for the total number of ducks in the dataset and confirms
that this is consistent with its observed data value. In the right
hand plot, the predictive probabilities have been first grouped into
centiles and then the observed proportion of ducks in each centile has
been plotted against the mid-point of each centile.  The plot shows
good agreement between the model predictions and observed proportions
as there is little deviation from the $45^o$ line.
Fig.~\ref{FigRunsGOF1} in the supplementary materials shows plots
similar to that in the right hand plot in Fig.~\ref{FigOvE} but gives a
more comprehensive picture. Instead of just showing the calibration of
duck predictions, this plot contains that for all numbers of runs
scored (grouped into intervals, typically of size ten).  Overall these
plots show that, although the model does not provide a perfect
calibration, it does give a fairly accurate description of runs
scored.

\section{Discussion}\label{sec:discussion}
The data clearly show that there is considerable within batsmen
variability in cricket scores and there is demonstrable evidence that
batsmen are especially vulnerable at the beginning of their innings.
Also the standard cricket batting average measure makes the
unreasonable assumption that run scoring follows a geometric
distribution.  Further the zero-inflated random effects Poisson model
(with log-linear factors) gives a good description of the runs scored
in Test matches. In terms of ranking players, we found that Sir Donald
Bradman, unsurprisingly, was the best player (under the model) and
there was relatively little uncertainty about his ranking. However,
there was considerable uncertainty in the rankings of players lower
down the list.

We compared our rankings with those of two established lists: one list
by career Test batting average and the other, the ICC Best-Ever Test
Championship Rating list. Not surprisingly we found disagreement
between all three lists. This illustrates a central problem in ranking
batsmen by a single number summary when there is a high level of
innings-to-innings variation in runs scored by each batsman. In these
circumstances it is more appropriate to summarise a career by a
distribution which recognises the uncertainty in these single number
summaries. In this paper we look at the player's overall ability
within a model which accounts for the high level of innings-to-innings
variation, various cricket-specific factors (not-outs, zero inflation)
and adjusts for various important player-independent factors.  Even
without such adjustments, simple data averages can easily mislead as
some batsmen play relatively few innings: the five number summary for
career innings played is $1-4-12-35-329$. We summarise our
understanding of the player's ability by a distribution or an interval
which accounts for uncertainty.  These summaries are impacted less by
circumstance and luck (such as when an lbw decision goes in the
batsman's favour and he makes a big score) than any fundamental
difference in ability.

The model represents the quality of the opposition through dynamic
opposition-specific parameters in order to capture potential changes
in the performance of Test playing countries over time, such as
periods of notable strength and weakness. Other factors were
considered for inclusion in the model but omitted due to data
limitations or in the interests of model parsimony. The effect of
playing at home was explicitly accounted for, and this could be
sub-categorised further into individual Test match grounds.  However,
although some grounds such as Lord's and the Sydney Cricket Ground
have been staples on the Test match roster, there has been a lot of
change in venues used in the sub-continent and so insufficient data is
available to be able to account for individual stadium effects. Test
matches are typically played as part of a series but data on the match
number within a series was not available in our dataset.  Similarly,
we might take account of the position of the batsmen in the batting
order. However, we believe that batting position is chosen to suit the
individual characteristics of each player in order to maximise runs
scored, and so leave out this factor from our model.

An obvious extension of this work would be to apply it to the
performance of both batsmen and bowlers. The approach could be further
extended to analyse data from one day internationals (ODIs), which,
despite only being an international sport since 1971, has already seen
around 3900 fixtures take place.  This equates to almost the same
amount of data as used here for Test matches since ODIs feature one
innings per team per match. A larger meta-model simultaneously
analysing the Test/ODI batsmen and bowlers could also help in
addressing the issue of ranking players as such a model would have the
potential to identify not only the quantity of the runs but also
refine attempts to ascertain their `quality' by explicitly factoring
in more granular data relating to the opposition, such as the strength
of the bowling attack in a given innings.  Twenty20 cricket is another
avenue for future work but currently there may be insufficient data
for an analysis of the type used in this work.

\bibliographystyle{apalike}
\bibliography{paper}

\appendix

\section{Model checking}

\begin{figure}[t!]
\centering
\includegraphics[width=\textwidth]{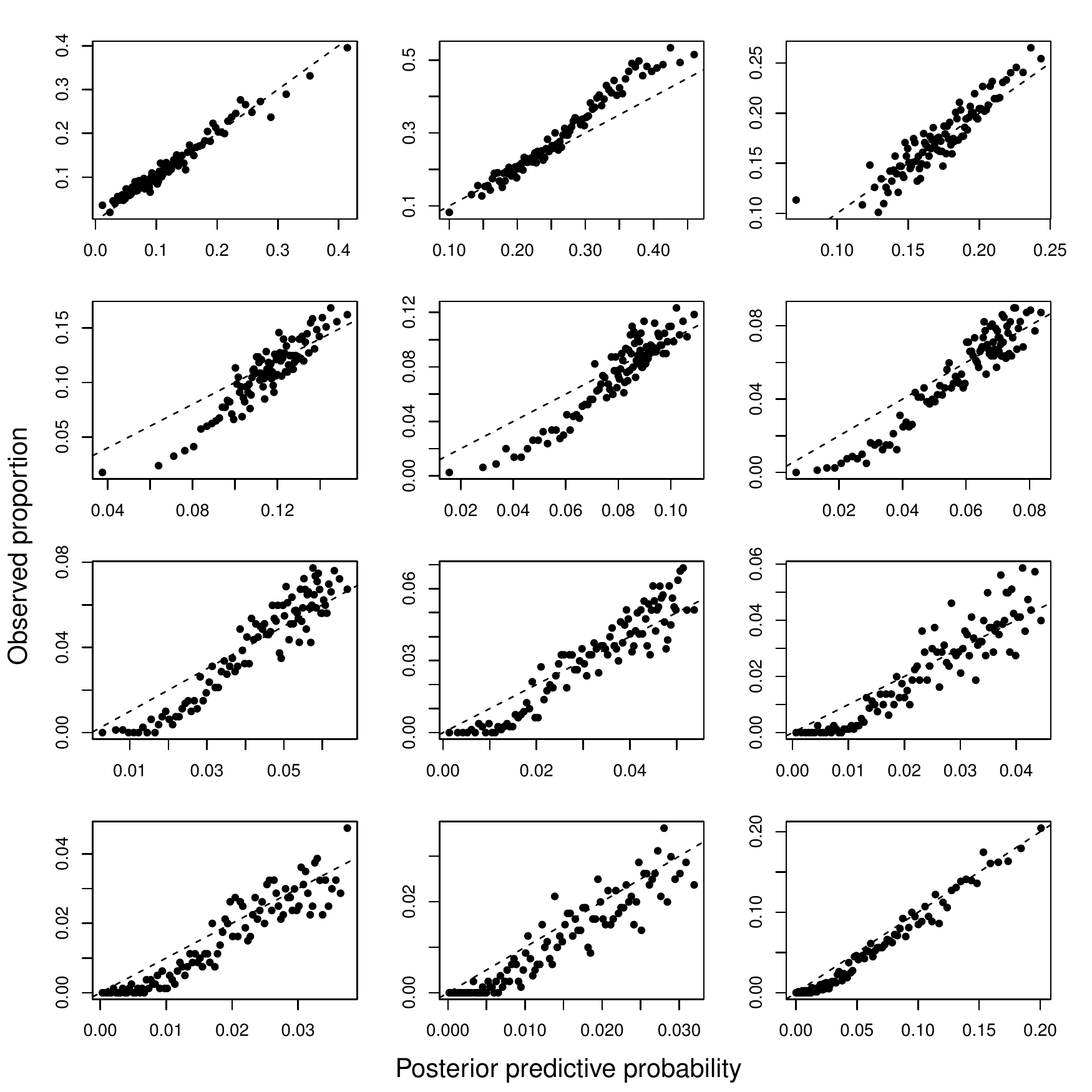}
\caption{Plots of observed proportion of runs scored against centiles
  of posterior predictive probabiities of runs scored organised by
  intervals of runs scored, reading from left-to-right: top row -- ducks, 1-9,
  10-19; second row -- 20-29, 30-39, 40-49; third row -- 50-59, 60-69, 70-79; bottom row -- 80-89, 90-99, centuries (100+)}
\label{FigRunsGOF1}
\end{figure}

Figure~\ref{FigRunsGOF1} shows similar model checking plots to that in
the left hand plot in Figure~\ref{FigOvE} in the paper but now gives a
more comprehensive picture. Instead of just showing the calibration of
duck predictions, this plot contains that for all numbers of runs
scored but, to save space, these have been grouped into intervals,
typically of size ten.  The top left plot gives again the calibration
of ducks for completeness. Note that the predictive distributions
typically have a long right tail due to the small values of
the~$\eta_i$.  Therefore, the range of predictive probabilities
encountered reduces as we look at plots for increasing numbers of
runs.  Overall these plots show that, although the model does not
provide a perfect calibration, it does give a fairly accurate
description of runs scored.

\section{Full conditionals and non-standard proposals}
The full conditional distributions and the proposal distributions used
in the MCMC scheme are as follows.

The posterior density can be factorised as
\begin{align*}
\pi(\kappavec,\etavec,\pivec\mid\xvec,\cvec,\dvec) 
&\propto 
\pi(\xvec,\cvec,\dvec\mid\lambdavec(\kappavec),\etavec,\pivec)\pi(\kappavec)\pi(\etavec) \pi(\pivec)\\
&\propto 
\pi(\xvec,\cvec,\dvec\mid\kappavec,\etavec,\pivec)\pi(\kappavec)\pi(\etavec) \pi(\pivec), 
\end{align*}
with $\lambdavec=\lambdavec(\kappavec)$, where $\xvec$, $\cvec$ and
$\dvec$ are the vectors of runs scored and associated censoring and
duck indicators respectively, and
$\kappavec=(\thetavec,\deltavec,\sigma_\delta,\alphavec,
\zeta_2,\nuvec,\xivec,\omegavec)$ contains the remaining parameters in
the model, with $\nuvec=(\nu_{2:4})$, $\xivec=(\xi_{2:10})$, and
$\omegavec=(\omega_{2:10,1:13})$. In the following derivations, some
expressions have been simplified by using
$\beta_{ijk}=\lambda_{ijk}/\eta_i$.

\subsubsection*{Update to the player effects}
The player effects $\theta_i$ are independent in the full conditional
distribution for~$\thetavec$. Therefore the MCMC scheme updates uses a
sweep of Metropolis-Hastings steps in which each $\theta_i$
$(i=1,\ldots,\numtestcricketers)$ is updated one at a time by using a
normal random walk proposal centred at the current value. The FCD for
$\theta_i$ is, for $\theta_i\in\mathbb{R}$
\begin{align*}
\pi(\theta_i \mid \cdot) 
&\propto \pi(\theta_i) \pi(\xvec,\cvec,\dvec\mid\kappavec,\etavec,\pivec)  \\
&\propto \exp\left\{-\frac{(\theta_i - \mu_{\theta_{d_i}})^2}{2\sigma_{\theta_{d_i}}^2} \right\}
\prod_{j,k}\left\{\pi_i+(1-\pi_i)(1+\beta_{ijk})^{-\eta_i}\right\}^{(1-c_{ijk})d_{ijk}}\\
&\qquad\times\prod_{j,k}\left[\left\{
\beta_{ijk}^{x_{ijk}}(1+\beta_{ijk})^{-(\eta_i+x_{ijk})}\right\}^{1-c_{ijk}} 
P(X\geq x_{ijk})^{c_{ijk}}\right]^{1-d_{ijk}}
\end{align*}

\begin{align*}
\phantom{\pi(\theta_i \mid \cdot)}
&\propto \exp\biggl\{-\frac{(\theta_i - \mu_{\theta_{d_i}})^2}{2\sigma_{\theta_{d_i}}^2} 
+\sum_{j,k} (1-c_{ijk})d_{ijk}\log\left\{\pi_i+(1-\pi_i)/(1+\lambda_{ijk}/\eta_i)^{\eta_i}\right\}
\biggr.\\
&\qquad\qquad+
\sum_{j,k}(1-c_{ijk})(1-d_{ijk})\{x_{ijk}\log\lambda_{ijk}-(\eta_i+x_{ijk})\log(1+\lambda_{ijk}/\eta_i)\}\\
&\biggl.\qquad\qquad\qquad\qquad+
\sum_{j,k}c_{ijk}(1-d_{ijk})\log P(X_{ijk}\geq x_{ijk})\biggr\}.
\end{align*}
Note that $X_{ijk}\sim NB\{\eta_i,\eta_i/(\eta_i+\lambda_{ijk})\}$.

\subsubsection*{Update to the year effects}
The posterior full conditional density (FCD) for the year effects is,
for $\deltavec\in\mathbb{R}^{139}$
\begin{align*}
\pi(\deltavec\mid\cdot) 
&\propto \pi(\deltavec|\sigma_\delta) \pi(\xvec,\cvec,\dvec \mid \kappavec,\etavec,\pivec)  \\
&\propto \exp\left\{-\frac{1}{2\sigma_\delta^2} \deltavec^TQ\deltavec\right\}
\prod_{i,j,k}\left\{\pi_i+(1-\pi_i)(1+\beta_{ijk})^{-\eta_i}\right\}^{(1-c_{ijk})d_{ijk}}\\
&\qquad\times\prod_{i,j,k}\left[\left\{
\beta_{ijk}^{x_{ijk}}(1+\beta_{ijk})^{-(\eta_i+x_{ijk})}\right\}^{1-c_{ijk}} 
P(X\geq x_{ijk})^{c_{ijk}}\right]^{1-d_{ijk}}\\
&\propto\exp\{-g(\deltavec)\},
\end{align*}
where
\begin{align*}
g(\deltavec)&=\sigma_\delta^{-2}\deltavec^TQ\deltavec/2
-\sum_{i,j,k} (1-c_{ijk})d_{ijk}\log\left\{\pi_i+(1-\pi_i)(1+\beta_{ijk})^{-\eta_i}\right\}
\biggr.\\
&\qquad\qquad-\sum_{i,j,k}(1-c_{ijk})(1-d_{ijk})\{x_{ijk}\log\lambda_{ijk}-(\eta_i+x_{ijk})\log(1+\beta_{ijk})\}\\
&\biggl.\qquad\qquad\qquad\qquad-
\sum_{i,j,k}c_{ijk}(1-d_{ijk})\log P(X_{ijk}\geq x_{ijk}).
\end{align*}
Using the second-order Taylor series expansion about $\tilde\deltavec$
\[
g(\deltavec) \simeq g(\tilde\deltavec) + g'(\tilde\deltavec)^T(\deltavec-\tilde\deltavec) + \frac{1}{2}(\deltavec-\tilde\deltavec)^T g''(\tilde\deltavec)(\deltavec-\tilde\deltavec)
\]
gives a normal approximation to the FCD, namely, 
\[
\deltavec\mid\cdot\overset{approx}{\sim }
N\left(\tilde\deltavec-g'(\tilde\deltavec)^Tg''(\tilde\deltavec)^{-1},~g''(\tilde\deltavec)^{-1}\right).
\]
In order to simplify the calculation of proposals, we ignore the
relatively small number of terms in $g(\deltavec)$ which use data on
ducks and not outs, that is, we use
\begin{align*}
g(\deltavec)&=\sigma_\delta^{-2}\deltavec^TQ\deltavec/2
-\sum_{i,j,k}(1-c_{ijk})(1-d_{ijk})\{x_{ijk}\log\lambda_{ijk}-(\eta_i+x_{ijk})\log(1+\beta_{ijk})\}.
\end{align*}
The derivative with respect to $\delta_\ell$ is
\begin{align*}
g(\deltavec)_\ell&=\sigma_\delta^{-2}(Q\deltavec)_\ell
-\sum_{i,j,k:y_{ijk}=\ell}(1-c_{ijk})(1-d_{ijk})\left\{\frac{x_{ijk}}{\lambda_{ijk}}-\frac{1+x_{ijk}/\eta_i}{1+\beta_{ijk}}\right\}\lambda_{ijk}\\
&=\sigma_\delta^{-2}(Q\deltavec)_\ell
-\sum_{i,j,k:y_{ijk}=\ell}(1-c_{ijk})(1-d_{ijk})\eta_i(x_{ijk}-\lambda_{ijk})/(\eta_i+\lambda_{ijk})
\end{align*}
and so the first and second order derviatives are
\[
g'(\deltavec)=\sigma_\delta^{-2}Q\deltavec+\bvec\qquad\text{and}\qquad
g''(\deltavec)=\sigma_\delta^{-2}Q-\diag(\cvec),
\]
where $\bvec=(b_\ell)$ and $\cvec_\ell=(c_{a\ell})$,
$b_{\ell}=\sum_{i,j,k:y_{ijk}=\ell}(1-c_{ijk})(1-d_{ijk})\eta_i(x_{ijk}+\eta_i)/(\eta_i+\lambda_{ijk})^2$
and
$c_\ell=2\sum_{i,j,k:y_{ijk}=\ell}(1-c_{ijk})(1-d_{ijk})\eta_i\lambda_{ijk}(x_{ijk}+\eta_i)/(\eta_i+\lambda_{ijk})^2$.
Therefore
\[
\deltavec\mid\cdot\overset{approx}{\sim } 
N\left(\tilde\deltavec-\{\sigma_\delta^{-2}Q\tilde\deltavec+\tilde\bvec\}^T
\{\sigma_\delta^{-2}Q-\diag(\tilde\cvec)\}^{-1},~\{\sigma_\delta^{-2}Q-\diag(\tilde\cvec)\}^{-1}\right),
\]
where $\tilde\bvec=\bvec(\tilde\deltavec)$ and
$\tilde\cvec=\cvec(\tilde\deltavec)$. This distribution could be used
to give proposals for $\deltavec$ about its current
value~$\tilde\deltavec$ in the MCMC scheme. However, as suggested by
\citet{RueGMRF}, we take $\tilde\deltavec$ as the mode of the FCD
(determined by iterating on the mean in this normal approximation) as
this leads to improved acceptance rates at the cost of increasing
computing time. We note that this leads to an independence proposal
distribution.

Also the update for the year effects smoothing parameter is conjugate,
with $\sigma_\delta^2|\cdot\sim
IG(a_\delta+139/2,b_\delta+\deltavec^TQ\deltavec/2)$.

\subsubsection*{Update to the player ageing parameters}
The ageing parameters $\alphavec_i=(\alpha_{1i},\alpha_{2i})$ are
independent in the full conditional distribution for~$\alphavec$.
Therefore the MCMC scheme updates $\alphavec$ using a sweep of
Metropolis-Hastings steps in which each $\alphavec_i$
$(i=1,\ldots,\numtestcricketers)$ is updated one at a time by using
independent normal and log-normal random walk proposals for
$\alpha_{1i}$ and $\alpha_{2i}$ respectively, each centred at the
current value.  The FCD for $\alphavec_i$ is, for
$\alpha_{1i}\in\mathbb{R}$, $\alpha_{2i}>0$
\begin{align*}
\pi(\alphavec_i\mid\cdot) 
&\propto \pi(\alphavec_i)\pi(\xvec,\cvec,\dvec\mid\kappavec,\etavec,\pivec)\\
&\propto\alpha_{2i}^{-1}
\exp\left\{-(\alpha_{1i}-30)^2/8-(\log\alpha_{2i}+3)^2/18\right\}
\times\prod_{j,k}\left\{\pi_i+(1-\pi_i)(1+\beta_{ijk})^{-\eta_i}\right\}^{(1-c_{ijk})d_{ijk}}\\
&\qquad\qquad\times\prod_{j,k}\left[\left\{
\beta_{ijk}^{x_{ijk}}(1+\beta_{ijk})^{-(\eta_i+x_{ijk})}\right\}^{1-c_{ijk}} 
P(X\geq x_{ijk})^{c_{ijk}}\right]^{1-d_{ijk}}\\
&\propto\alpha_{2i}^{-1}
\exp\left\{-(\alpha_{1i}-30)^2/8-(\log\alpha_{2i}+3)^2/18\right.\\
&\qquad\quad+\sum_{j,k}(1-c_{ijk})d_{ijk}\log\left\{\pi_i+(1-\pi_i)/(1+\lambda_{ijk}/\eta_i)^{\eta_i}\right\}\\
&\qquad\qquad\qquad+\sum_{j,k}(1-c_{ijk})(1-d_{ijk})\{x_{ijk}\log\lambda_{ijk}-(\eta_i+x_{ijk})\log(1+\lambda_{ijk}/\eta_i)\}\\
&\qquad\qquad\qquad\qquad\left.+\sum_{j,k}c_{ijk}(1-d_{ijk})\log P(X_{ijk}\geq x_{ijk})\right\}.
\end{align*}

\subsubsection*{Update to the playing away from home effect}
The MCMC scheme updates the playing away from home effect $\zeta_2$
using a Metropolis-Hastings step and a normal random walk proposal
centred at the current value. The FCD for $\zeta_2$ is, for
$\zeta_2\in\mathbb{R}$
\begin{align*}
\pi(\zeta_2\mid\cdot) 
&\propto \pi(\zeta_2)\pi(\xvec,\cvec,\dvec\mid\kappavec,\etavec,\pivec)\\
&\propto \exp\left(-2\zeta_2^2\right)\prod_{i,j,k:h_{ijk}=2}\left\{\pi_i+(1-\pi_i)(1+\beta_{ijk})^{-\eta_i}\right\}^{(1-c_{ijk})d_{ijk}}\\
&\qquad\times\prod_{i,j,k:h_{ijk}=2}\left[\left\{
\beta_{ijk}^{x_{ijk}}(1+\beta_{ijk})^{-(\eta_i+x_{ijk})}\right\}^{1-c_{ijk}} 
P(X\geq x_{ijk})^{c_{ijk}}\right]^{1-d_{ijk}}\\
&\propto \exp\left\{-2\zeta_2^2+\sum_{i,j,k:h_{ijk}=2}(1-c_{ijk})d_{ijk}\log\left\{\pi_i+(1-\pi_i)/(1+\lambda_{ijk}/\eta_i)^{\eta_i}\right\}\right.\\
&\qquad\qquad\qquad+\sum_{i,j,k:h_{ijk}=2}(1-c_{ijk})(1-d_{ijk})\{x_{ijk}\log\lambda_{ijk}-(\eta_i+x_{ijk})\log(1+\lambda_{ijk}/\eta_i)\}\\
&\qquad\qquad\qquad\qquad\left.+\sum_{i,j,k:h_{ijk}=2}c_{ijk}(1-d_{ijk})\log P(X_{ijk}\geq x_{ijk})\right\}.
\end{align*}


\subsubsection*{Update to the innings effects}
The MCMC scheme updates the innings effects $\nuvec$ using a sweep of
Metropolis-Hastings steps in which each $\nu_g$ $(g=2,3,4)$ is updated
one at a time by using a normal random walk proposal centred at the
current value. The FCD for $\nu_g$ is, for $\nu_g\in\mathbb{R}$
\begin{align*}
\pi(\nu_g\mid\cdot) 
&\propto \pi(\nu_g)\pi(\xvec,\cvec,\dvec\mid\kappavec,\etavec,\pivec)\\
&\propto \exp\left(-2\nu_g^2\right)\prod_{i,j,k:m_{ijk}=g}\left\{\pi_i+(1-\pi_i)(1+\beta_{ijk})^{-\eta_i}\right\}^{(1-c_{ijk})d_{ijk}}\\
&\qquad\times\prod_{i,j,k:m_{ijk}=g}\left[\left\{
\beta_{ijk}^{x_{ijk}}(1+\beta_{ijk})^{-(\eta_i+x_{ijk})}\right\}^{1-c_{ijk}} 
P(X\geq x_{ijk})^{c_{ijk}}\right]^{1-d_{ijk}}\\
&\propto \exp\left\{-2\nu_g^2+\sum_{i,j,k:m_{ijk}=g}(1-c_{ijk})d_{ijk}\log\left\{\pi_i+(1-\pi_i)/(1+\lambda_{ijk}/\eta_i)^{\eta_i}\right\}\right.\\
&\qquad\qquad\qquad+\sum_{i,j,k:m_{ijk}=g}(1-c_{ijk})(1-d_{ijk})\{x_{ijk}\log\lambda_{ijk}-(\eta_i+x_{ijk})\log(1+\lambda_{ijk}/\eta_i)\}\\
&\qquad\qquad\qquad\qquad\left.+\sum_{i,j,k:m_{ijk}=g}c_{ijk}(1-d_{ijk})\log P(X_{ijk}\geq x_{ijk})\right\}.
\end{align*}


\subsubsection*{Update to the opposition effects}
The MCMC scheme updates the opposition effects $\xivec$ using a
sweep of Metropolis-Hastings steps in which each $\xi_q$
$(q=2,\ldots,10)$ is updated one at a time by using a normal random
walk proposal centred at the current value. The FCD for $\xi_q$ is,
for $\xi_q\in\mathbb{R}$
\begin{align*}
\pi(\xi_q\mid\cdot) 
&\propto \pi(\xi_q)\pi(\xvec,\cvec,\dvec\mid\kappavec,\etavec,\pivec)\\
&\propto \exp\left(-2\xi_q^2\right)\prod_{i,j,k:o_{ijk}=q}\left\{\pi_i+(1-\pi_i)(1+\beta_{ijk})^{-\eta_i}\right\}^{(1-c_{ijk})d_{ijk}}\\
&\qquad\times\prod_{i,j,k:o_{ijk}=q}\left[\left\{
\beta_{ijk}^{x_{ijk}}(1+\beta_{ijk})^{-(\eta_i+x_{ijk})}\right\}^{1-c_{ijk}} 
P(X\geq x_{ijk})^{c_{ijk}}\right]^{1-d_{ijk}}\\
&\propto \exp\left\{-2\xi_q^2+\sum_{i,j,k:o_{ijk}=q}(1-c_{ijk})d_{ijk}\log\left\{\pi_i+(1-\pi_i/(1+\lambda_{ijk}/\eta_i)^{\eta_i}\right\}\right.\\
&\qquad\qquad\qquad+\sum_{i,j,k:o_{ijk}=q}(1-c_{ijk})(1-d_{ijk})\{x_{ijk}\log\lambda_{ijk}-(\eta_i+x_{ijk})\log(1+\lambda_{ijk}/\eta_i)\}\\
&\qquad\qquad\qquad\qquad\left.+\sum_{i,j,k:o_{ijk}=q}c_{ijk}(1-d_{ijk})\log P(X_{ijk}\geq x_{ijk})\right\}.
\end{align*}

\newpage\subsubsection*{Update to the opposition$\times$decade interaction effects}
The MCMC scheme updates the interaction effects $\omegavec$ using a
sweep of Metropolis-Hastings steps in which each $\omega_{qd}$
$(q = 2,\ldots,10; d=1,\ldots,13)$ is updated one at a time by using a normal random
walk proposal centred at the current value. The FCD for $\omega_{qd}$ is,
for $\omega_{qd}\in\mathbb{R}$
\begin{align*}
\pi(\omega_{qd}\mid\cdot) 
&\propto \pi(\omega_{qd})\pi(\xvec,\cvec,\dvec\mid\kappavec,\etavec,\pivec)\\
&\propto \exp\left(-2\omega_{qd}^2\right)\prod_{i,j,k:o_{ijk} = q, e_{ijk}=d}\left\{\pi_i+(1-\pi_i)(1+\beta_{ijk})^{-\eta_i}\right\}^{(1-c_{ijk})d_{ijk}}\\
&\qquad\times\prod_{i,j,k:o_{ijk} = q, e_{ijk}=d}\left[\left\{
\beta_{ijk}^{x_{ijk}}(1+\beta_{ijk})^{-(\eta_i+x_{ijk})}\right\}^{1-c_{ijk}} 
P(X\geq x_{ijk})^{c_{ijk}}\right]^{1-d_{ijk}}\\
&\propto \exp\left\{-2\omega_{qd}^2+\sum_{i,j,k:o_{ijk} = q, e_{ijk}=d}(1-c_{ijk})d_{ijk}\log\left\{\pi_i+(1-\pi_i)/(1+\lambda_{ijk}/\eta_i)^{\eta_i}\right\}\right.\\
&\qquad\qquad\qquad+\sum_{i,j,k:o_{ijk} = q, e_{ijk}=d}(1-c_{ijk})(1-d_{ijk})\{x_{ijk}\log\lambda_{ijk}-(\eta_i+x_{ijk})\log(1+\lambda_{ijk}/\eta_i)\}\\
&\qquad\qquad\qquad\qquad\left.+\sum_{i,j,k:o_{ijk} = q, e_{ijk}=d}c_{ijk}(1-d_{ijk})\log P(X_{ijk}\geq x_{ijk})\right\}.
\end{align*}

\subsubsection*{Update to the random effects heterogeneity parameters}
The random effect parameters $\eta_i$ are independent in the full
conditional distribution for~$\etavec$. Therefore the MCMC scheme
updates uses a sweep of Metropolis-Hastings steps in which each
$\eta_i$ $(i=1,\ldots,\numtestcricketers)$ is updated one at a time by using a
normal random walk proposal centred at the current value on the log
scale. The FCD for $\eta_i$ is, for $\eta_i>0$
\begin{align*}
\pi(\eta_i\mid\cdot) 
& \propto \pi(\eta_i)\pi(\xvec,\cvec,\dvec\mid\kappavec,\etavec,\pivec)\\
& \propto \eta_i^{-1} \exp\left\{-\frac{(\log\eta_i)^2}{2}\right\} 
\prod_{j,k}\left\{\pi_i+(1-\pi_i)(1+\beta_{ijk})^{-\eta_i}\right\}^{(1-c_{ijk})d_{ijk}}\\
&\qquad\times\prod_{j,k}\left[(1-\pi_i)\left\{\binom{x_{ijk}+\eta_i-1}{x_{ijk}}
\beta_{ijk}^{x_{ijk}}(1+\beta_{ijk})^{-(\eta_i+x_{ijk})}\right\}^{1-c_{ijk}} 
P(X\geq x_{ijk})^{c_{ijk}}\right]^{1-d_{ijk}}\\
&\propto \exp\left\{-\log\eta_i-\frac{(\log\eta_i)^2}{2}+\sum_{j,k}(1-c_{ijk})d_{ijk}\log\left\{\pi_i+(1-\pi_i)/(1+\lambda_{ijk}/\eta_i)^{\eta_i}\right\}\right.\\
&\quad\qquad\qquad+\sum_{j,k}(1-c_{ijk})(1-d_{ijk})\left\{-x_{ijk}\log\eta_i-(\eta_i+x_{ijk})\log(1+\lambda_{ijk}/\eta_i)+\log\binom{x_{ijk}+\eta_i-1}{x_{ijk}}\right\}\\
&\qquad\qquad\qquad\qquad\left.+\sum_{j,k}c_{ijk}(1-d_{ijk})\log P(X_{ijk}\geq x_{ijk})\right\}.
\end{align*}

\subsubsection*{Update to the zero-inflation parameters}
The zero-inflation parameters $\pi_i$ are independent in the full
conditional distribution for~$\pivec$. Therefore the MCMC scheme
updates uses a sweep of Metropolis-Hastings steps in which each
$\pi_i$ $(i=1,\ldots,\numtestcricketers)$ is updated one at a time by
using a normal random walk proposal centred at the current value on
the logit scale. The FCD for $\pi_i$ is, for $0<\pi_i<1$
\begin{align*}
\pi(\pi_i\mid\cdot) & \propto \pi(\pi_i) \pi(\xvec,\cvec,\dvec\mid\kappavec,\etavec,\pivec)\\
& \propto (1-\pi_i)^8\prod_{j,k}\left\{\pi_i+(1-\pi_i)(1+\beta_{ijk})^{-\eta_i}\right\}^{(1-c_{ijk})d_{ijk}}\\
&\qquad\times\prod_{j,k}\left[(1-\pi_i)\left\{\binom{x_{ijk}+\eta_i-1}{x_{ijk}}
\beta_{ijk}^{x_{ijk}}(1+\beta_{ijk})^{-(\eta_i+x_{ijk})}\right\}^{1-c_{ijk}} 
P(X\geq x_{ijk})^{c_{ijk}}\right]^{1-d_{ijk}}\\
&\propto (1-\pi_i)^{8+\sum_{j,k}(1-c_{ijk})(1-d_{ijk})}\prod_{j,k}\left\{\pi_i+(1-\pi_i)/(1+\lambda_{ijk}/\eta_i)^{\eta_i}\right\}^{(1-c_{ijk})d_{ijk}}.
\end{align*}

\subsubsection*{Update to the player-specific ability hyperparameters}
The MCMC scheme updates the player-specific hyperparameters $\mu_\theta$ and $\sigma_\theta$ using Gibbs steps, with
\begin{align*}
\mu_\theta\mid\cdot&\sim N\left(\frac{m_\mu \sigma_\theta^2 + \numtestcricketers\bar{\theta}s_\mu^2}{\sigma_\theta^2 + \numtestcricketers s_\mu^2},\frac{s_\mu^2\sigma_\theta^2}{\sigma_\theta^2 + \numtestcricketers s_\mu^2}\right)
\quad\text{and}\quad
\sigma_\theta^2\mid\cdot\sim IG\left(a_\sigma + \frac{\numtestcricketers}{2}, b_\sigma + \frac{1}{2}\sum_{i=1}^{\numtestcricketers}(\theta_i - \mu_\theta)^2\right).
\end{align*}

\section{Glossary of cricketing terms}
\hang\textbf{Batting average:} the average number of runs scored per
innings by a batsman, calculated by dividing the batsman's total runs
scored during those innings in question by the number of times the
batsman was out

\hang\textbf{Boundary:} the perimeter of the ground 

\hang\textbf{Century:} an individual batsman's score of at least 100 runs

\hang\textbf{Declaration:} the act of a captain voluntarily bringing
his side's innings to a close, in the belief that their score is now
great enough to prevent defeat

\hang\textbf{Dismissal:} to get one of the batsmen out so that he must
cease batting

\hang\textbf{Duck:} a batsman's score of nought (zero) dismissed, as in
`he was out for a duck'

\hang\textbf{Follow-on:} where the team that bats second is forced to
take its second batting innings immediately after its first, because
the team was not able to get close enough to the score achieved by the
first team batting in the first innings

\hang\textbf{Four:} a shot that reaches the boundary after touching the
ground, scoring four runs for the batting side

\hang\textbf{Innings:} one player's or one team's turn to bat (or bowl)

\hang\textbf{Leg before wicket (lbw):} a way of dismissing the batsman.
In brief, the batsman is out if, in the opinion of the umpire, the
ball hits any part of the batsman's body (usually the leg) before
hitting the bat and would have gone on to hit the stumps

\hang\textbf{Not out:} a batsman who is in and has not yet been dismissed 

\hang\textbf{Over:} the delivery of six consecutive legal balls by one bowler 

\hang\textbf{Partnership:} the number of runs scored between a pair of
batsmen before one of them gets dismissed

\hang\textbf{Run:} a term used in cricket for the basic means of
scoring. A single run is scored when a batsman has hit the ball with
his/her bat and are able to run the length (22 yards) of the pitch
before the ball is returned. The batsman may run more than once and
each completed run increments the scores of both the team and the
batsman

\hang\textbf{Six:} a shot which passes over or touches the boundary
without having bounced or rolled, so called because it scores six runs
to the batting side

\hang\textbf{Stumps:} the three vertical posts making up the wicket

\hang\textbf{Test match:} a cricket match with play spread over five
days with unlimited overs played between two senior international
teams. Considered to be the highest level of the game

\hang\textbf{Wicket:} the dismissal of a batsman or a set of stumps

\end{document}